\documentstyle[12pt,epsf]{article}
\newcommand{\abseta}{\mid \eta \mid \leq}
\def \bbar {\overline b}
\def \pt {p_{\rm T}}
\def \Kt {k_{\rm T}}
\def \Et {{\rm E}_{\rm T}}
\def \deg {^\circ}
\def \muprob {{\cal P}(p_{\rm T}^{\mu},\pt^b)}
\def \etprob {{\cal P}({\rm E}_{\rm T}^{\bbar},\pt^{\bbar})}
\def \pjet {P_{jet}}
\def\r#1{\ignorespaces $^{#1}$}
\begin{document}
\baselineskip 18pt

\begin{flushright}
Fermilab-Pub-95/289-E \\
\end{flushright}
\begin{center}
{\Large {\bf Measurement of correlated $\mu - \bbar$ jet cross
sections in  $p{\overline p}$ collisions at $\sqrt{s}=1.8$ TeV}}
\end{center}

\font\eightit=cmti8
\hfilneg
\begin{sloppypar}
\noindent
F.~Abe,\r {13} M.~G.~Albrow,\r 7 S.~R.~Amendolia,\r {23}
D.~Amidei,\r {16}
J.~Antos,\r {28} C.~Anway-Wiese,\r 4 G.~Apollinari,\r {26}
H.~Areti,\r 7
M.~Atac,\r 7 P.~Auchincloss,\r {25}
% ...
\end{sloppypar}
\vskip .025in
\begin{center}
(CDF Collaboration)
\end{center}

\vskip .025in
\begin{center}
\r 1  {\eightit Argonne National Laboratory, Argonne, Illinois 60439}
\\
\r 2  {\eightit Istituto Nazionale di Fisica Nucleare, University of
Bologna, I-40126 Bologna, Italy} \\
\r 3  {\eightit Brandeis University, Waltham, Massachusetts 02254} \\
\r 4  {\eightit University of California at Los Angeles, Los
Angeles, California  90024} \\
\r 5  {\eightit University of Chicago, Chicago, Illinois 60637} \\
\r 6  {\eightit Duke University, Durham, North Carolina  27708} \\
\r 7  {\eightit Fermi National Accelerator Laboratory, Batavia,
Illinois  60510} \\
\r 8  {\eightit Laboratori Nazionali di Frascati, Istituto Nazionale
di Fisica Nucleare, I-00044 Frascati, Italy} \\
\r 9  {\eightit Harvard University, Cambridge, Massachusetts 02138}
\\
\r {10} {\eightit University of Illinois, Urbana, Illinois 61801} \\
\r {11} {\eightit Institute of Particle Physics, McGill University,
Montreal  H3A 2T8, and University of Toronto,\\ Toronto M5S 1A7,
Canada} \\
\r {12} {\eightit The Johns Hopkins University, Baltimore, Maryland
21218} \\

\r {13} {\eightit National Laboratory for High Energy Physics (KEK),
Tsukuba,  Ibaraki 305, Japan} \\
\r {14} {\eightit Lawrence Berkeley Laboratory, Berkeley, California
94720} \\
\r {15} {\eightit Massachusetts Institute of Technology, Cambridge,
Massachusetts  02139} \\
\r {16} {\eightit University of Michigan, Ann Arbor, Michigan 48109}
\\
\r {17} {\eightit Michigan State University, East Lansing, Michigan
48824} \\
\r {18} {\eightit University of New Mexico, Albuquerque, New Mexico
87131} \\
\r {19} {\eightit Osaka City University, Osaka 588, Japan} \\
\r {20} {\eightit Universita di Padova, Istituto Nazionale di Fisica
Nucleare, Sezione di Padova, I-35131 Padova, Italy} \\
\r {21} {\eightit University of Pennsylvania, Philadelphia,
Pennsylvania 19104} \\
\r {22} {\eightit University of Pittsburgh, Pittsburgh, Pennsylvania
15260} \\
\r {23} {\eightit Istituto Nazionale di Fisica Nucleare, University
and Scuola Normale Superiore of Pisa, I-56100 Pisa, Italy} \\
\r {24} {\eightit Purdue University, West Lafayette, Indiana 47907}
\\
\r {25} {\eightit University of Rochester, Rochester, New York 14627}
\\
\r {26} {\eightit Rockefeller University, New York, New York 10021}
\\
\r {27} {\eightit Rutgers University, Piscataway, New Jersey 08854}
\\
\r {28} {\eightit Academia Sinica, Taiwan 11529, Republic of China}
\\
\r {29} {\eightit Superconducting Super Collider Laboratory, Dallas,
Texas 75237} \\
\r {30} {\eightit Texas A\&M University, College Station, Texas
77843} \\
\r {31} {\eightit Texas Tech University, Lubbock, Texas 79409} \\
\r {32} {\eightit University of Tsukuba, Tsukuba, Ibaraki 305, Japan}
\\
\r {33} {\eightit Tufts University, Medford, Massachusetts 02155} \\
\r {34} {\eightit University of Wisconsin, Madison, Wisconsin 53706}
\\
\r {35} {\eightit Yale University, New Haven, Connecticut 06511} \\

\vspace{0.3in}

We report on measurements of differential $\mu - \bbar$ cross
sections, where the muon is from a semi-leptonic $b$ decay and  the
$\bbar$ is identified using precision track reconstruction in jets.
The semi-differential correlated cross sections,
d$\sigma$/d$\Et^{\bbar}$, d$\sigma$/d$\pt^{\bbar}$, and
d$\sigma$/d$\delta\phi(\mu - \bbar)$ for $\pt^{\mu}>$~9~GeV/c,
$|\eta^{\mu}|<$~0.6, $\Et^{\bbar}>$~10~GeV,  $|\eta^{\bbar}|<$~1.5,
are presented and compared to next-to-leading order QCD
calculations.\\
\end{center}

\section{Introduction}
\label{section-introduction}

\par Measurements of $b$ production in $p{\overline p}$ collisions
provide quantitative test of perturbative QCD.  Single integral $b$
cross section measurements at $\sqrt{s}$ = 1.8 TeV have been
systematically higher than predictions from next-to-leading order
(NLO) QCD calculations~\cite{cdf_bmu,D0-mu,FMNR}.  These cross
section measurements, from inclusive $b  \rightarrow$ lepton decays
and exclusive $B$ meson decays ($B^+\rightarrow$J/$\psi K^+$),
 use the kinematical relationship between the decay product
(e.g, the lepton) and the $b$ quark spectra to obtain the production
cross section integrated over a rapidity range $\mid y \mid <$ 1 and
a $\pt$ range from a threshold $\pt^{min}$ to infinity.  Single
differential $B$ meson cross section measurements~\cite{CDF_Bmes} are
also systematically higher than the NLO prediction.

\par Semi-differential $b - \bbar$ cross sections give further
information on the underlying QCD production mechanisms by exploring
the kinematical correlations between the two $b$ quarks.  Comparison
of NLO predictions with experimental measurements can give
information on whether higher order corrections serve as a scale
factor to the NLO prediction or change the production distributions.
As future high precision $B$ decay measurements at hadron colliders
($e.g.$, CP violation studies in $B^0\rightarrow$J/$\psi
K^0_s$~\cite{snowmass_beta}) may depend upon efficient identification
of the decay products of both $b$ quarks, understanding of the
correlated cross sections is necessary.

\par This paper describes measurements of $\mu - \bbar$ correlated
cross sections as a function of the  jet transverse  energy
(d$\sigma$/d$\Et$, where $\Et = E \times sin \theta$) and  transverse
momentum (d$\sigma$/d$\pt$) of the $\bbar$ and as a function of the
azimuthal separation (d$\sigma$/d$\delta\phi$) between the muon and
$\bbar$ jet, for $\pt^{\mu}> 9$ GeV/c, $|\eta^{\mu}|<$ 0.6,
$\Et^{\bbar}>$ 10 GeV/c, $|\eta^{\bbar}|<$ 1.5.  The data are $15.08
\pm 0.54$ $\rm{pb^{-1}}$ of $p\overline{p}$ collisions at  $\sqrt{s}
= 1.8$ TeV collected with the CDF detector between  August, 1992 and
May, 1993.  We make use of two features of $B$ hadrons to separate
them from the large jet backgrounds at 1.8 TeV: the high branching
fraction into muons ($\approx$ 10\%~\cite{PDG}) and the  relatively
long lifetime ($\approx$ 1.5 picoseconds~\cite{PDG}).  The advent of
precision silicon microstrip detectors, with hit resolutions
approaching 15 $\mu$m, provides  the ability to efficiently identify
the hadronic decays  of $B$ hadrons as well as the semi-leptonic
decays.

\par We use the identification of a high transverse momentum muon as
the initial signature of the presence of $b$ quarks.  In
$p\overline{p}$ collisions, high transverse momentum muons come from
the production and decay of heavy quarks ($c,b,t$), vector bosons
($W,Z\deg$), and light mesons ($\pi,K$).  Additional identification
techniques are necessary to convert a $\mu -$ jet cross section into
a $\mu - \bbar$ cross section.

\par For these measurements, the first $b$ is identified from a
semi-leptonic decay muon and the other $b$ (referred to for
simplicity as the $\bbar$, though we do not perform explicit flavor
identification for either $b$) is identified by using precision track
reconstruction in jets to measure  the displaced particles from
$\bbar$ decay.  Jets are identified as clusters of energy in the
calorimeter~\cite{jet_papers}.  In this paper, a jet energy (or jet
transverse energy) refers to the measured energy in the cluster.   A
procedure to simultaneously unfold the effects of detector response
and resolution is used to translate the results from $\bbar$ jets to
$\bbar$ quarks.

\par  It should be noted that we have chosen to report the
measurements as differential $\mu - \bbar$ cross sections rather than
$b - \bbar$ cross sections in order to facilitate comparison to
calculations of the production cross sections.  The process of
converting a muon cross section to a quark cross section includes
systematic uncertainties~\cite{cdf_bmu} with strong dependence on
both production, fragmentation, and decay models.  By presenting $\mu
- \bbar$ cross sections, we facilitate the future comparison of the
experimental results to different models, since the data results and
uncertainties are not tied to specific models.

\par Section~\ref{section-detector} describes the detector systems
used for muon and $\bbar$ jet identification.
Section~\ref{section-dataset} contains descriptions of the muon and
jet identification requirements.  The $\bbar$ jet counting is
discussed in section~\ref{section-correlated}.  In
section~\ref{section-acceptance}, the muon and $\bbar$ jet
identification efficiencies and acceptances are described.  The cross
section results, the calculation of additional physics backgrounds,
and jet to quark unfolding  are discussed in
section~\ref{section-results}.   Section~\ref{section-conclusions}
closes with a discussion of the experimental results.

\section{Detector Description}

\label{section-detector}

\par The CDF has been described in detail elsewhere~\cite{NIM_book}.
The analysis presented in this paper depends on the tracking and muon
systems for triggering and selection, while identification of
hadronic jets  uses the information from the calorimeter elements.

\subsection{Tracking and Muon Systems}

\par This analysis uses the silicon vertex detector
(SVX)~\cite{SVXNIM},  the vertex drift chamber (VTX) and the central
tracking chamber (CTC)~\cite{CTCNIM} for charged particle tracking.
These are all located in a 1.4 T solenoidal magnetic field.  The SVX
consists of 4 layers of silicon-strip detectors with $r-\phi$
readout, including pulse height information, with a total active
length of 51 cm in the range $-27.3 < z < 27.3$
cm~\cite{coordinates}.  The pitch between readout strips is 60 $\mu$m
on the inner 3 layers and 55 $\mu$m on the outermost layer. A single
point spatial resolution of 13 $\mu$m has been obtained.  The first
measurement plane is located 2.9 cm from the interaction point,
leading to an impact parameter resolution of $\approx$ 15 $\mu$m for
tracks with transverse momentum, $\pt$, greater than 5 GeV/c.  The
VTX is a  time projection chamber providing information out to a
radius of 22 cm and $\mid \eta \mid <$ 3.5.  The VTX is used to
measure the $p{\overline p}$ interaction  vertex ($z_0$) along the
$z$ axis with a resolution of 1 mm. The CTC is a cylindrical drift
chamber containing 84 layers, which are grouped into alternating
axial and stereo superlayers containing 12 and 6 wires respectively,
covering the radial range from 28 cm to 132 cm.  The momentum
resolution of the CTC is $\delta$$\pt$/$\pt$ = 0.002~$\times~\pt$
for isolated tracks (where $\pt$ is in GeV/c).  For tracks found in
both the SVX and CTC, the momentum resolution improves to
0.0009~$\times~\pt \oplus 0.0066$ (where $\pt$ is in GeV/c).

\par  The muon system consists of two detector elements.  The Central
Muon system (CMU)~\cite{CMUNIM}, which consists of four layers of
limited streamer chambers located at a radius of 384 cm, behind
$\approx$ 5 absorption lengths of material, provides muon
identification for the pseudorapidity range $|\eta| < $0.6.  This
$\eta$ region is further instrumented by the Central Muon upgrade
system (CMP)~\cite{CMP}, which is a set of four chambers located
after $\approx$ 8 absorption lengths of material.  Approximately 84\%
of the solid angle of  $|\eta|\leq$0.6 is covered by the CMU, 63\% by
the CMP, and 53\% by both.  Muon transverse momentum is measured with
the charged tracking systems and has the tracking resolutions
described above.  CMU (and CMP) segments are defined as a set of 2 or
more hits along radially aligned wires.

\subsection{Calorimeter Systems}

\par This analysis uses the CDF central and plug calorimeters, which
are segmented into separate electromagnetic and hadronic
compartments.  In all cases, the absorber in the electromagnetic
compartment is lead, and in the hadronic compartment, iron. The
central region subtends the range $|\eta|< $ 1.1 and spans 2$\pi$ in
azimuthal coverage, with scintillator as the active medium.  The plug
region subtends the range $1.1 < |\eta| < 2.4$ with gas proportional
chambers as the active media, again  with 2$\pi$ azimuthal coverage.
The calorimeters have resolutions that range from 13.7\%/$\sqrt{{\rm
E}_{\rm T}} \oplus 2\%$ for the central electromagnetic to
106\%/$\sqrt{{\rm E}_{\rm T}} \oplus 6\%$ for the plug
hadronic~\cite{TOPPRD}.

\subsection{Trigger System}

\par  CDF uses a three-level trigger system~\cite{TRIGGERNIM}.  Each
level is a logical OR of a number of triggers designed to select
events with electrons, muons or jets.  The analysis presented in this
paper uses only the muon trigger path.  Section~\ref{section-dataset}
includes a description of the trigger efficiencies for muons.

\par The Level 1 central muon trigger requires a pair of hits on
radially aligned wires in the CMU system.  The $\pt$ of the track
segment is measured using the arrival times of the drift electrons at
the wires to determine the deflection angle due to the magnetic
field.  The trigger requires that the segment have $\pt>$ 6 GeV/c,
with at least two confirming hits in the projecting CMP  chambers.

\par The Level 2 trigger includes information from a list of $r-\phi$
tracks found by the central fast tracker (CFT)~\cite{CFTNIM}, a
hardware track processor which uses fast timing information from the
CTC as input.  The CFT momentum resolution is
$\delta\pt$/$\pt~\approx 0.035 \times \pt$, with a plateau efficiency
of 91.3$\pm$0.3\% for tracks with $\pt$ above 12 GeV/c.   The CMU
chamber segment is required to match a CFT track with $\pt >$ 9.2
GeV/c within 5$\deg$ in the $\phi$ coordinate.

\par The Level 3 trigger makes use of a slightly modified version of
the offline software reconstruction algorithms, including full 3
dimensional track reconstruction.  The CMU segment is required to
match a CTC track with $\pt>$ 7.5 GeV/c, extrapolated to the chamber
radius, within 10 cm in $r-\phi$.  Confirming CMP hits are required.

\section{Dataset Selection} \label{section-dataset}

\par Beginning with the sample of muon triggered events, we select
events with both a well identified muon candidate and a minimum
transverse energy jet.  A primary vertex is found by a weighted fit
of the VTX $z_0$ vertex position and SVX tracks.  An iterative search
removes tracks with large impact parameters (the distance of closest
approach in the $r - \phi$ plane) from the fit. Since the $\bbar$ jet
identification technique (described in
section~\ref{section-correlated}) depends upon the precision track
reconstruction in the SVX, we require the event primary vertex $\mid
z_0 \mid < 30$ cm. In this section, we discuss the identification
variables, efficiency, and geometric acceptance for muon and $\bbar$
jet candidates.  Table~\ref{table-efficiencies} contains a summary of
the muon efficiency and acceptance results and
table~\ref{table-bbar_jet_acceptance} contains a summary of the
$\bbar$ jet identification and acceptance results.

\subsection{Muon Identification}

\par  Muons are identified as a well matched coincidence between a
track in the CTC and segments in both the CMU and CMP muon systems.
The CTC track is required to have $\pt > 9$ GeV/c and point back to
within 5 cm in $z$ of the found primary vertex.   The measured track
is extrapolated to the muon chambers and is required to match the
muon chamber track segment position to $<~3\sigma$ in the transverse
direction (for both CMU and CMP) and $<\sqrt{12}\sigma$ in the
longitudinal direction (for CMU).  In all cases, $\sigma$ includes
the contributions from smearing due to multiple scattering in the
absorber and the muon chamber resolution.  We require that the track
be found in the SVX.

\par
\begin{figure}
\epsfysize=7in
\epsffile[0 90 594 684]{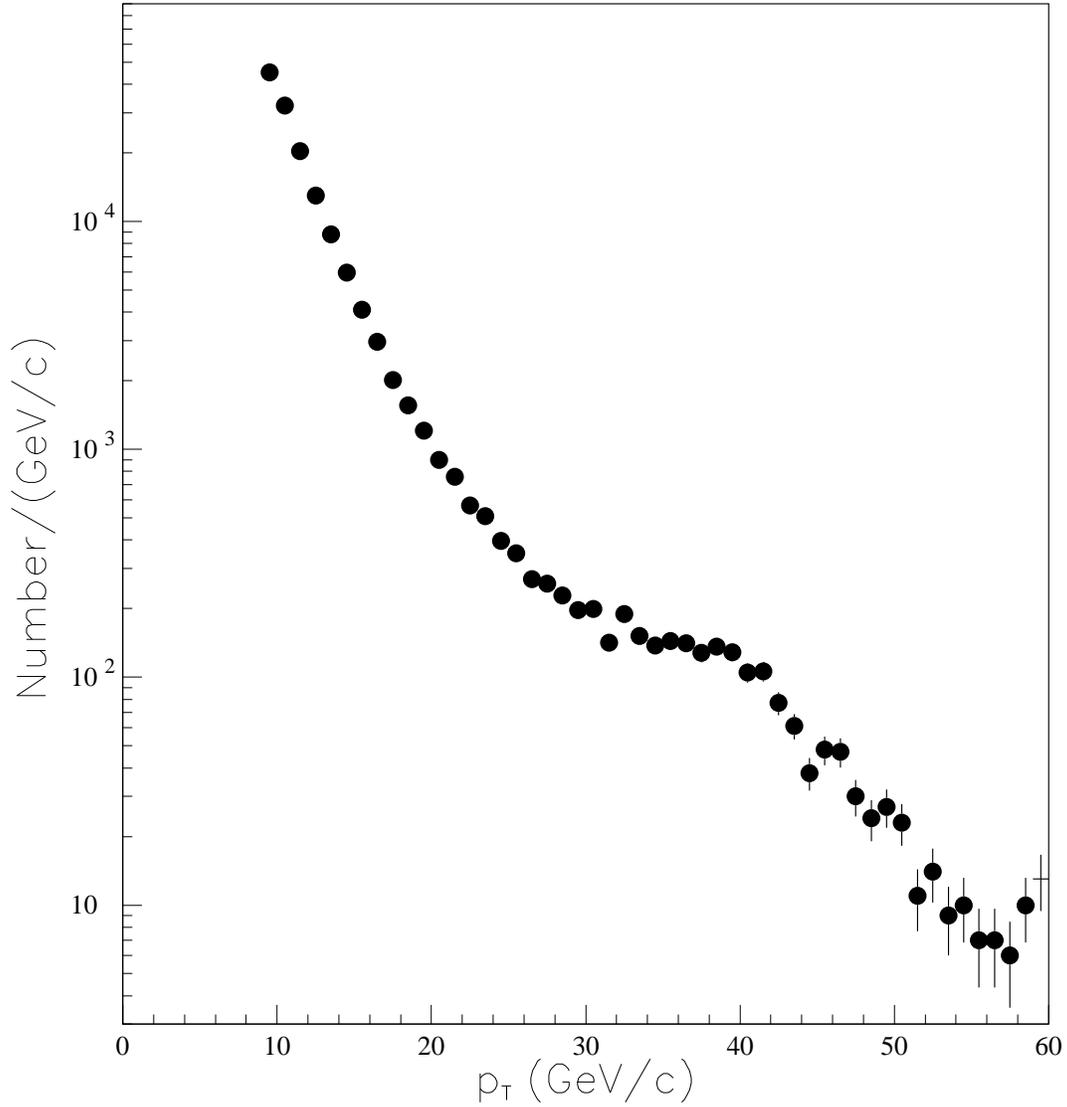}
\caption{The $\mu$ $\pt$ spectrum for the 9 GeV/c sample.  There are
144097 events, with 80 having $\pt >$ 60 GeV/c.  The enhancement
above 25 GeV/c is due to the presence of muons from W and Z boson
decays.}
\label{fig-muon_pt_spectrum}
\end{figure}

 There are 144097 events passing all the muon requirements in
this data sample.  In the case where there is more than 1 identified
muon in an event, we take the highest $\pt$ muon as the $b$ candidate
muon.   The fraction of muons from $b$ decay is measured to be
approximately 40\%~\cite{TOPPRD}, with a fraction from charm decays
of approximately 20\%.  Figure~\ref{fig-muon_pt_spectrum} shows the
transverse momentum spectrum for the muons in this dataset.  The
flattening of the  slope at high $\pt$ is due to muons from
electroweak boson decay.

\subsection{Jet Identification}

\par Jets are identified in the CDF calorimeter systems using a fixed
cone (in $\eta - \phi$ space) algorithm.  A detailed description of
the algorithm can be found in reference~\cite{jet_papers}.  For this
analysis, we use a cone radius of 0.4.  We require that jets have
transverse energy, $\Et$ = E $\times$ sin~$\theta$ (where $E$ is the
total energy in the cone), greater than 10 GeV, and $\abseta$ 1.5.
There are 50154 events passing the muon and jet $\Et$ requirements.
We use tracking techniques to identify $\bbar$ jets, so the
pseudorapidity range is restricted to the region with tracking
coverage.  All jet energies in this paper are measured energies, not
including corrections for known detector effects({\em e.g.},
calorimeter non-linearities).  An unsmearing procedure, described in
section~\ref{section-results}, is used to convert  measured jet $\Et$
distributions to parton momentum distributions.

\par We associate SVX tracks to a jet by requiring that the track be
within the cone of 0.4 around the jet axis.  To remove tracks
consistent with photon conversions and $K_S$ or $\Lambda$ decays
originating from the primary vertex, we require that the impact
parameter, $d$, be less than 0.15 cm.  In addition, track pairs
consistent with $K_S \rightarrow \pi^+\pi^-$ or $\Lambda \rightarrow
p\pi$ decays are removed.  We select jets with two or more well
measured tracks~\cite{TOPPRD}, $\pt >$ 1 GeV/c, with positive impact
parameters. The impact parameter sign is defined to be $+1$ for
tracks where the point of closest approach to the primary vertex lies
in the  same hemisphere as the jet direction, and $-1$ otherwise.
for 

\par  We require that the distance, $\Delta$R, in $\eta - \phi$ space
between the muon and the jet axis  be greater than 1.0.  There are
16842 events passing all the muon and jet requirements.  The
$\Delta$R separation is chosen so that tracks clustered around the
jet axis are separated from the muon direction, in order to have
physical separation of the $b$ and $\bbar$ decay products.  As there
may be more than one jet in an event passing these requirements, we
select the jet with the lowest jet probability (defined in
section~\ref{section-correlated}), so as to have a unique combination
of $\mu$ --- jet in each event.

\section{$\bbar$ Jet Counting}
\label{section-correlated}

\par The $\bbar$ jet is not identified on an event-by-event basis,
but instead by fitting for the number of $\bbar$ jets present in the
sample.  For each jet, we combine the impact parameter information
for tracks in the jet cone into one number which describes the
probability that the given collection of tracks has no decay products
from long lived particles.  In a $\bbar$ jet,  there will be a
significant number of tracks from the $B$ hadron decay, and hence the
probability for a $\bbar$ jet will be much less than 1.

\subsection{The Jet Probability Algorithm}

\par The $\bbar$ jet identification makes use of a probability
algorithm~\cite{ALEPH}  which compares track impact parameters to
measured resolution functions in order to calculate for each jet a
probability that there are no long lived particles in the jet cone.
This probability is uniformly distributed for light quark or gluon
jets (we refer to these jets as prompt jets), but is very low for
jets with displaced vertices from heavy flavor decay.   We now
briefly describe the transformation from the track impact parameters
to the jet probability measure.

\par
\begin{figure}
\epsfysize=7in
\epsffile[0 90 594 684]{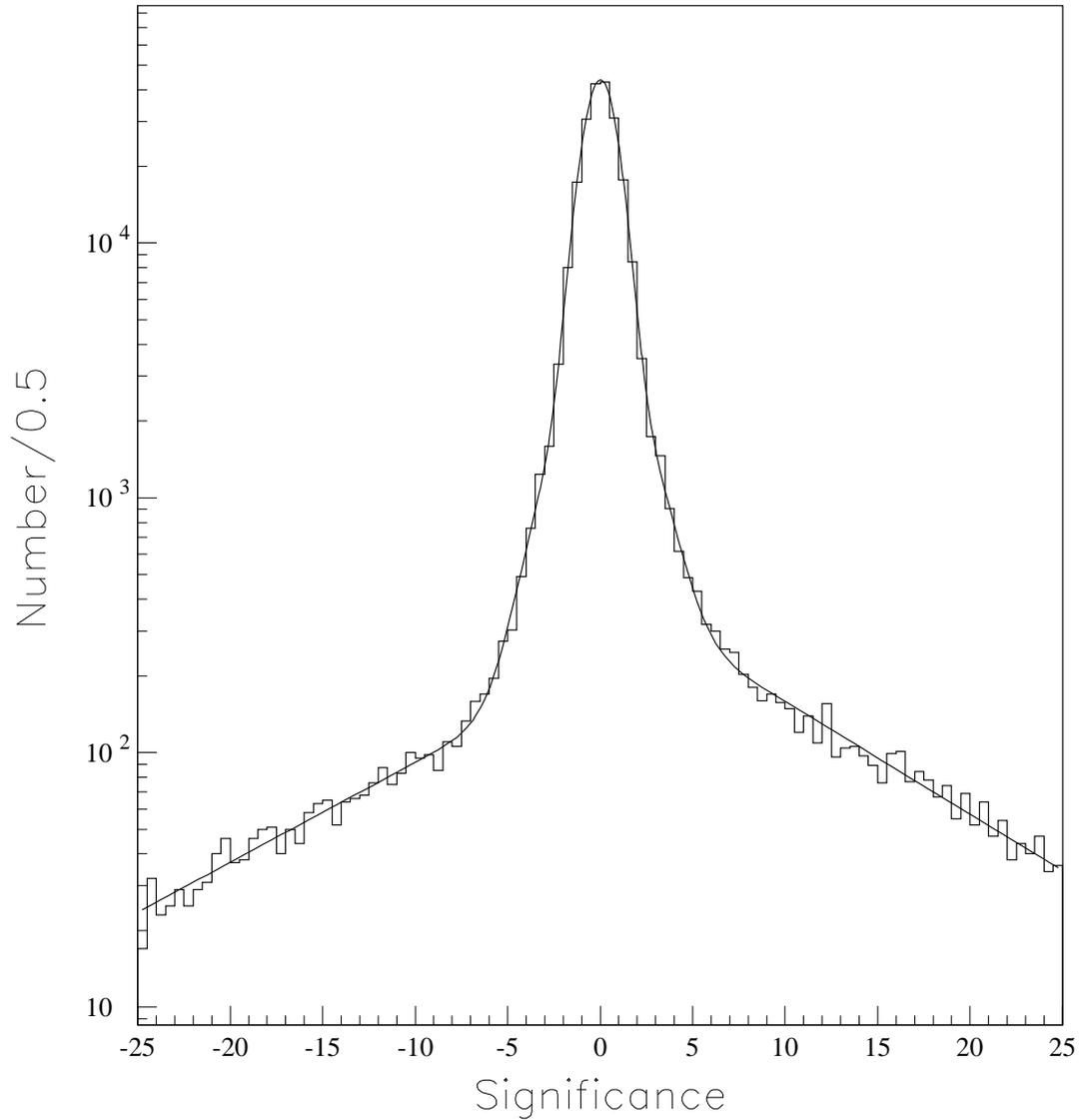}
\caption{A sample track resolution function, including fits to both
positive and negative signed impact parameters.  The function is fit
to 2 gaussians plus two exponentials, one for the positive side and
one for the negative side.  The excess on the positive side is
attributable to long lived particles in the sample.}
\label{figure-resolution_function}
\end{figure}

 The track impact parameter significance is defined as the value
of the impact parameter divided by the uncertainty in that quantity,
which includes both the measured uncertainties from the track and
primary vertex reconstruction.
Figure~\ref{figure-resolution_function} shows the distribution of
impact parameter significance ($s_0 = d/\sigma$) from a sample of
jets taken with a 50 GeV jet trigger~\cite{TOPPRD}, overlayed with a
fitted function.  The tails of the distribution come from a
combination of non-Gaussian effects and true long lived particles.
Using a combination of data and Monte Carlo simulation of heavy
flavor decays, we estimate approximately 30\% of the tracks with
$\mid s_0 \mid >$ 3.0 are from the decay products of long lived
particles, which is consistent with the excess in the positive $s$
side of the distribution. The negative side of the fitted function,
$R(s)$, is used to  map the impact parameter significance $s_0$  to a
track probability measure:

\begin{equation}
P(s_0) = \frac{\int_{-\infty}^{-|s_0|}R(s)ds}
{\int_{-\infty}^{0}R(s)ds}.
\end{equation}

\noindent  The track probability is a measure of the probability of
getting a track with impact parameter significance greater than $s$.
The function $R(s)$ can be defined for both Monte Carlo simulated
datasets and the jet dataset.  The mapping of the resolution function
to the track probability distribution removes differences in the
resolution between the simulated detector performance and the true
detector performance and creates a variable which is consistent
between the two datasets.

\par   The jet probability is then calculated from the independent
track probabilities as: \begin{equation} P_{jet} =
\Pi\sum_{k=0}^{N-1}\frac{(-\ln\Pi)^k}{k!}, \end{equation} where
\begin{equation} \Pi = P_1 P_2 \cdots P_N \end{equation} \noindent is
the product of the individual probabilities of the selected tracks.
For the rest of this paper, when the track selection requirements
pick tracks with negative signed impact parameters, we will refer to
the measure as the ``negative jet probability''.  When the track
selection requirements pick tracks with positive signed impact
parameters, we will refer to the measure as $P_{jet}$.

\par
\begin{figure}
\epsfysize=7in
\epsffile[0 90 594 684]{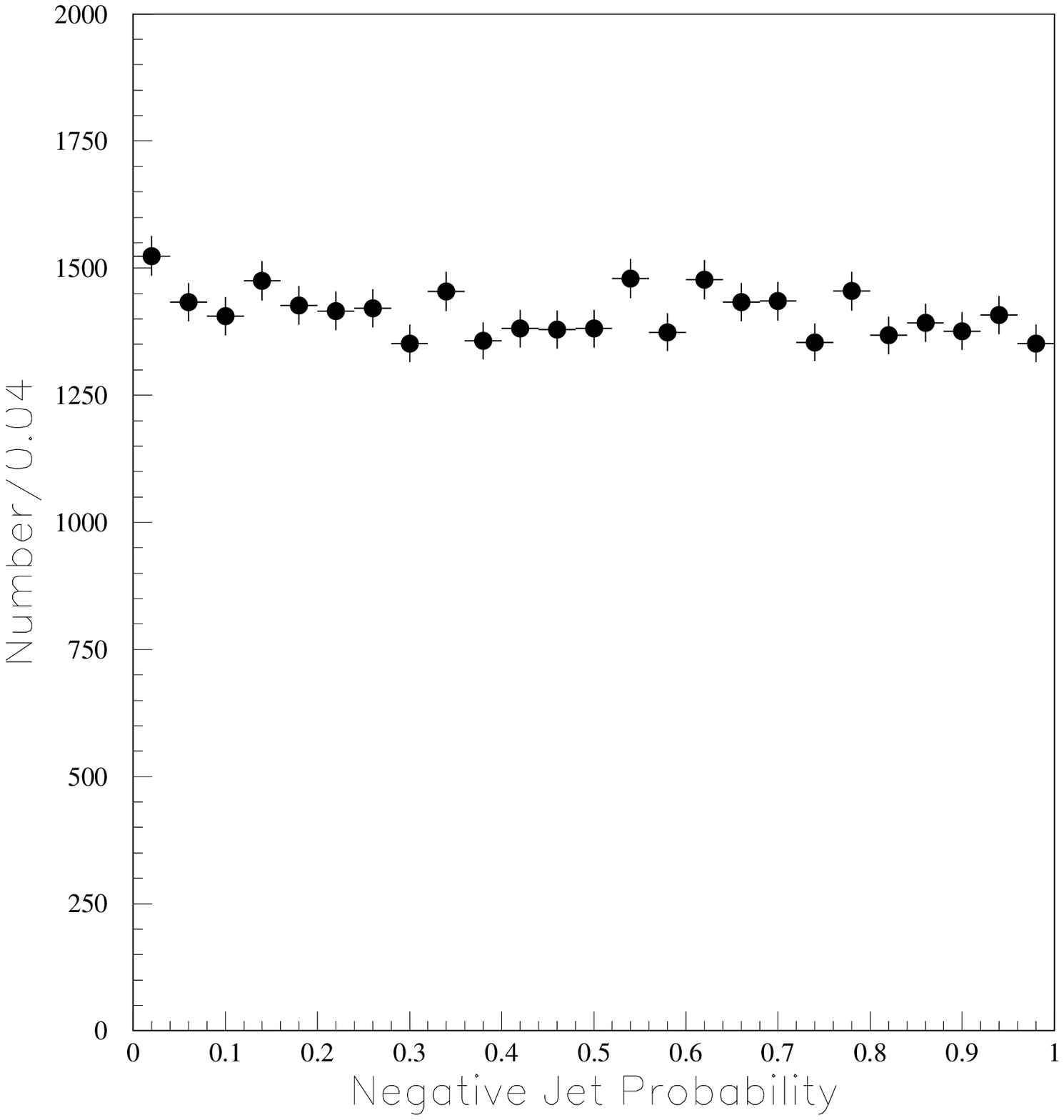}
\caption{The negative jet probability spectrum, calculated using
tracks with negative signed impact parameters, in a sample of 50 GeV
jets.}
\label{figure-jet_prob_samples}
\end{figure}

 Figure~\ref{figure-jet_prob_samples} shows the distribution of
the negative jet probability in the 50 GeV sample.  Since this
distribution reflects the smearing of the impact parameter
significance distribution due to resolution effects, we expect that
the $\pjet$ distribution for prompt (light quark and gluon) jets to
be similar.  Simulated jets containing heavy flavor decays show
distinct differences from this distribution, peaking at low values of
$\pjet$.  In figure~\ref{fig-smoothed_shapes}, we show the
distributions of log$_{10}(\pjet)$ for $\bbar$, charm, and prompt
jets.

\par We have found that the $\pjet$ shape for heavy flavor jets is
affected by the number of tracks used in the calculation of $\pjet$
which are also used in the primary vertex fit.  The turnover visible
in the $\bbar$ and charm distributions around -3 in log$_{10}(\pjet)$
is a combination of the vertex requirements ($d/\sigma <$ 3 for
tracks in the fit) and the $\bbar$ and charm lifetimes.  $\bbar$ and
charm jets are affected differently, due to differences in lifetime
and decay multiplicities.

\subsection{$\bbar$ Jet Fit Technique}

\par
\begin{figure}
\epsfysize=7in
\epsffile[0 90 594 684]{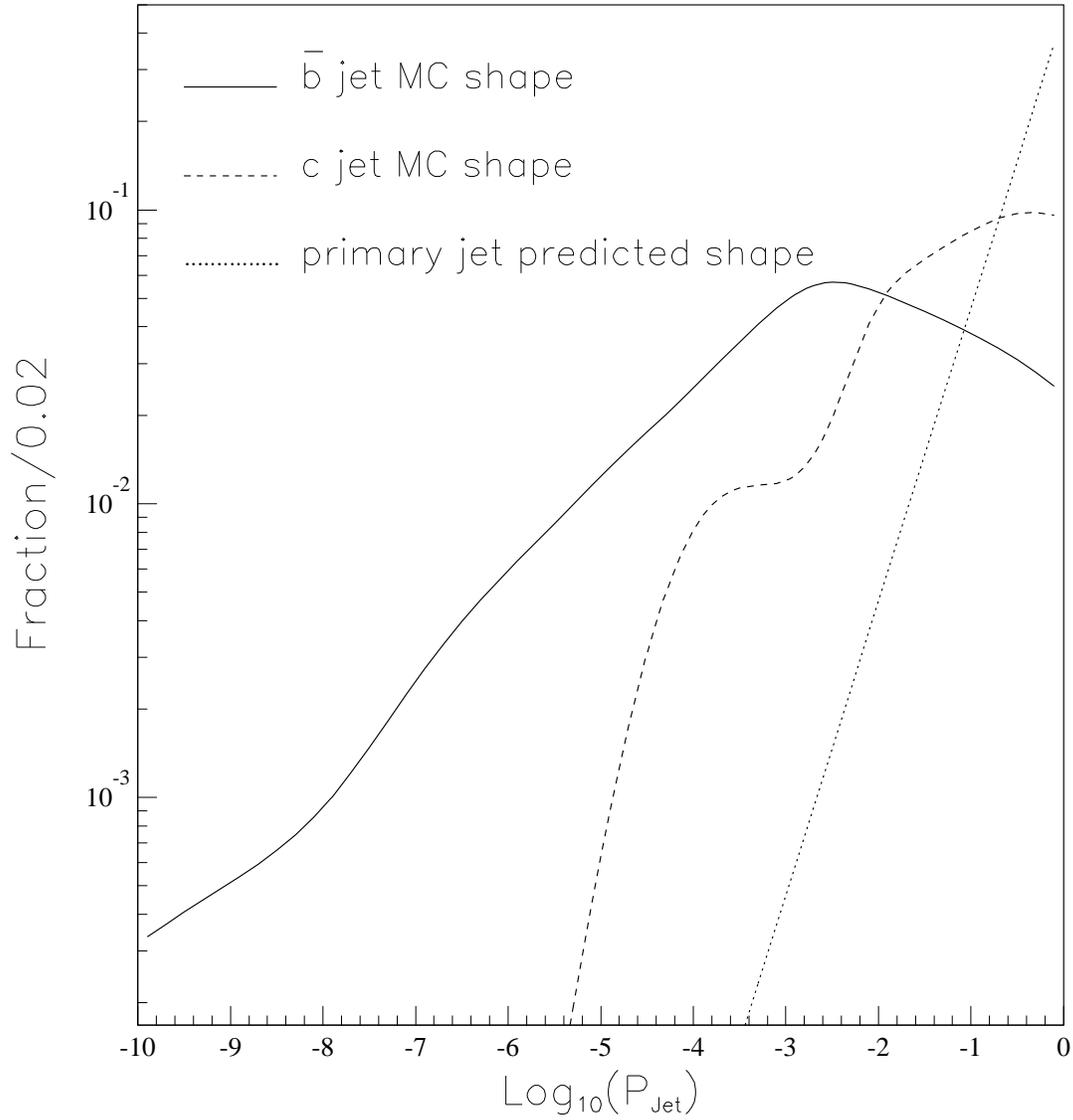}
\caption{The log$_{10}(\pjet)$ distributions used as inputs to the
fitting program.  The $b$ and $c$ shapes are smoothed versions of
Monte Carlo distributions, while the primary shape is an exponential
function.  The three distributions are normalized to equal area and
shown on the same vertical scale.}
\label{fig-smoothed_shapes}
\end{figure}

 We use a binned maximum likelihood fit to distinguish the
$\bbar$, $c$, and prompt jet contributions in the sample. For a
binned likelihood fit, we find that log$_{10}(\pjet)$ shows stronger
differentiation between $\bbar$, $c$, and prompt jets (see
figure~\ref{fig-smoothed_shapes}) than $\pjet$  and use this variable
in the fitting algorithm. We fit over the  range -10 --- 0 in
log$_{10}(\pjet)$, where the $\bbar$, $c$, and prompt contributions
are constrained to be positive. No other constraints are included in
the fit.  We model the prompt jets with an exponential  distribution,
since a logarithm transforms a uniform distribution to an
exponential distribution.

\par We have explored the effect of different Monte Carlo samples to
construct the  input shape used in the fit.  Using different input
$\bbar$ jet Monte Carlo samples (see
section~\ref{section-acceptance}) compared to the test distribution
shows a 5\% change in the fit fractions.  Changing the average
$\bbar$ lifetime by 6\%~\cite{b_lifetime,lifetime_choice} changed the
fit fraction by 3\%.  We include a 5.8\% systematic uncertainty to
our fit results to account for systematic uncertainties in the
fitting procedure and uncertainty on the $\bbar$ lifetime.

\begin{figure}
\epsfysize=7in
\epsffile[0 90 594 684]{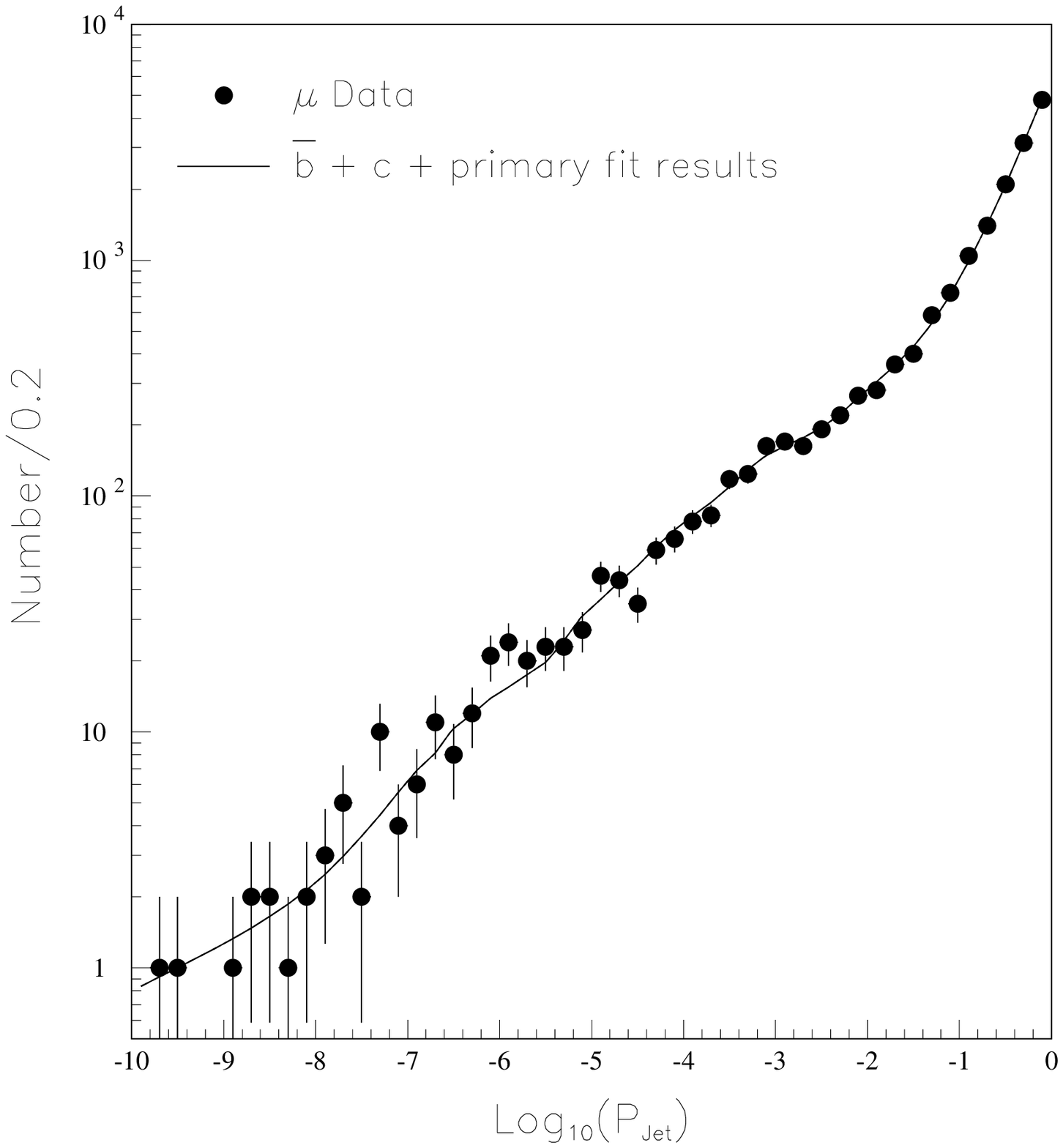}
\caption{For all jets ($\Et >$ 10 GeV) in the $\mu$ sample, we show
the data distribution  overlayed with the fit results.  There are two
events in the data with log$_{10}(\pjet) <$ -10.  Statistical errors
on the data and the fit results are included.  The fit results model
the data well over the entire range of the fit.}
\label{fig-sample_fit}
\end{figure}

\par
\begin{figure}
\epsfysize=7in
\epsffile[0 90 594 684]{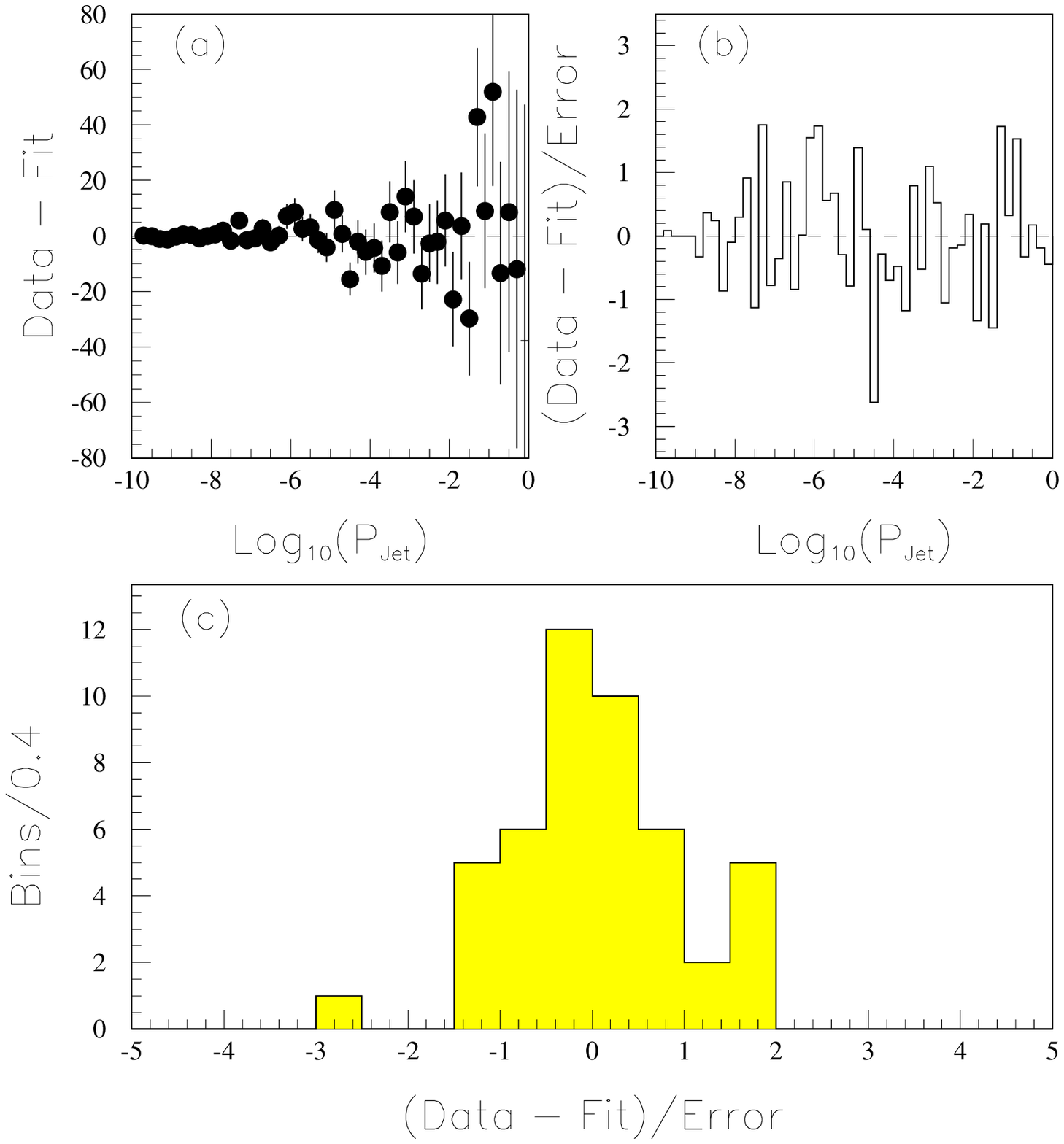}
\caption{Various comparisons of the data distribution and the fit
results.  We show (a) the bin by bin difference between the data and
the fit results, (b) the bin-by-bin difference scaled to the errors,
and (c) the distribution of the difference scaled to the errors, with
mean 0.04 and RMS = 0.95.  In all cases, the errors are the
statistical error in the data points and the fitted results.}
\label{fig-fit_comparisons}
\end{figure}

  In figure~\ref{fig-sample_fit}, we show the distribution of
log$_{10}(\pjet)$ for all jets, $\Et >$ 10 GeV, in the muon sample,
overlayed with  the fit results.   In this sample, the fit finds 2484
$\pm$ 94 $\bbar$ jets, 1988 $\pm$ 175 $c$ jets, and 12368 $\pm$ 157
prompt jets for a total of 16840.  There are 16842 events in the data
sample.  Figure~\ref{fig-fit_comparisons} shows three comparisons of
the data and fit results, showing the bin-by-bin difference in the
results, the bin-by-bin difference divided by the errors, and the
distribution of the difference divided by the errors.  In these
distributions, the errors are  the statistical errors in the data
points. We do not include any error on the Monte Carlo shapes. From
these distributions, we can see that the inputs model the data well.
The difference divided by the errors has a mean of 0.04 and RMS of
0.95.

\par For the semi-differential measurements, we do an independent fit
of the log$_{10}(\pjet)$ distribution and then correct for the
acceptance in each $\Et$ or $\delta\phi$ bin.
Table~\ref{table-fit_results} contains a summary of the number of
total jets and the number of $\bbar$ jets in each $\Et$ and
$\delta\phi$ bin considered.

\begin{table}[tb]
\begin{center}
\begin{tabular}{|c|c|c|}       \hline
$\Et$ Range & Number of Jets & Estimated Number  \\
& & of $\bbar$ Jets \\ \hline\hline
10 --- 15 & 5174 & 547 $\pm$ 49 \\
15 --- 20 & 3818 & 618 $\pm$ 47 \\
20 --- 25 & 2563 & 453 $\pm$ 39 \\
25 --- 30 & 1698 & 278 $\pm$ 30 \\
30 --- 40 & 1921 & 327 $\pm$ 33 \\
40 --- 50 &  819 & 140 $\pm$ 20 \\
50 --- 100 & 849 & 107 $\pm$ 19 \\  \hline \hline
$\delta\phi$ Range &  &    \\ \hline\hline
0---$\frac{\pi}{8}$ & 43 & 4.8 $+5.5 - 4.8$ \\
$\frac{\pi}{8}$ --- $\frac{\pi}{4}$ & 83 & 25.0 $\pm$ 8.6 \\
$\frac{\pi}{4}$ --- $\frac{3\pi}{8}$ & 230 & 54.7 $\pm$ 13.3 \\
$\frac{3\pi}{8}$ --- $\frac{\pi}{2}$ & 336 & 78.2 $\pm$ 15.9 \\
$\frac{\pi}{2}$ --- $\frac{5\pi}{8}$ & 519 & 105. $\pm$ 18.5 \\
$\frac{5\pi}{8}$ --- $\frac{3\pi}{4}$ & 1008 & 160. $\pm$ 25. \\
$\frac{3\pi}{4}$ --- $\frac{7\pi}{8}$ & 3229 & 461. $\pm$ 42. \\
$\frac{7\pi}{8}$ --- $\pi$ & 11394 & 1593. $\pm$ 75. \\
\hline
\end{tabular}
\end{center}
\caption{$\bbar$ fit results as a function of jet $\Et$ and
$\delta\phi$ between the muon and $\bbar$ jet.  We have not included
a common systematic uncertainty of 5.8\%.}
\label{table-fit_results}
\end{table}

\section{Acceptance and Efficiency}
\label{section-acceptance}

\subsection{Muon Requirements}

\par  The muon geometric acceptance is the fraction of events with a
muon in the good fiducial region of the CMU and CMP chambers,
starting from a sample where the muon has $\pt >$ 9 GeV/c and $\mid
\eta \mid <$ 0.6.  Note that this term is only a geometric acceptance
and does not include kinematical cuts on the muon.

\par  The geometric acceptance is studied with a $b\rightarrow\mu$
Monte Carlo generator (which includes the sequential decays
$b\rightarrow c \mu$), with the input spectra coming from the next to
leading order  calculation of $b - \bbar$ production by Mangano,
Nason, and Ridolfi (MNR)~\cite{MNR}.  The input spectra use the MRSD0
structure functions~\cite{MRSD0} and renormalization scale $\mu_0 =
\sqrt{m_b^2 + (\pt^{b 2} + \pt^{\bbar 2}2)/2}$, with $m_b$ = 4.75
GeV/$c^2$. This generator produces $b$ quarks and $B$ hadrons, using
the Peterson fragmentation form~\cite{Peterson} with
$\epsilon=0.006~\pm~0.002$~\cite{fragmentation}.  $B$ hadrons are
decayed according to the CLEO Monte Carlo program, QQ~\cite{CLEOMC}.
We select events with a $b \rightarrow \mu$ decay, with muon $\pt >$
9 GeV/c and $\mid \eta \mid <$ 0.6.

\par For these studies, event vertices are distributed along the $z$
axis as a Gaussian with mean = -1.4 cm and $\sigma$ = 26.65
cm~\cite{Badgett_thesis}, which is a good approximation to the
average conditions seen in the data.  The muons are propagated to the
CMU and CMP chamber radii, including the effects of the central
magnetic field and multiple scattering.  The acceptance is then
defined as the fraction of muons which are in the good fiducial area
of both the CMU and CMP chambers and is found to be 53.0 $\pm$ 0.3\%
(statistical), independent of variations of the $\epsilon$ parameter
from 0.004 to 0.008.

\par  The muon trigger and selection depends significantly upon the
track reconstruction efficiency in the CTC.  We have defined our
efficiencies to be multiplicative, so that we can measure them
independently.  In this section, the efficiencies of the individual
selection requirements, and methods of  measuring them, are
described.

\par
\begin{figure}
\epsfysize=7in
\epsffile[0 90 594 684]{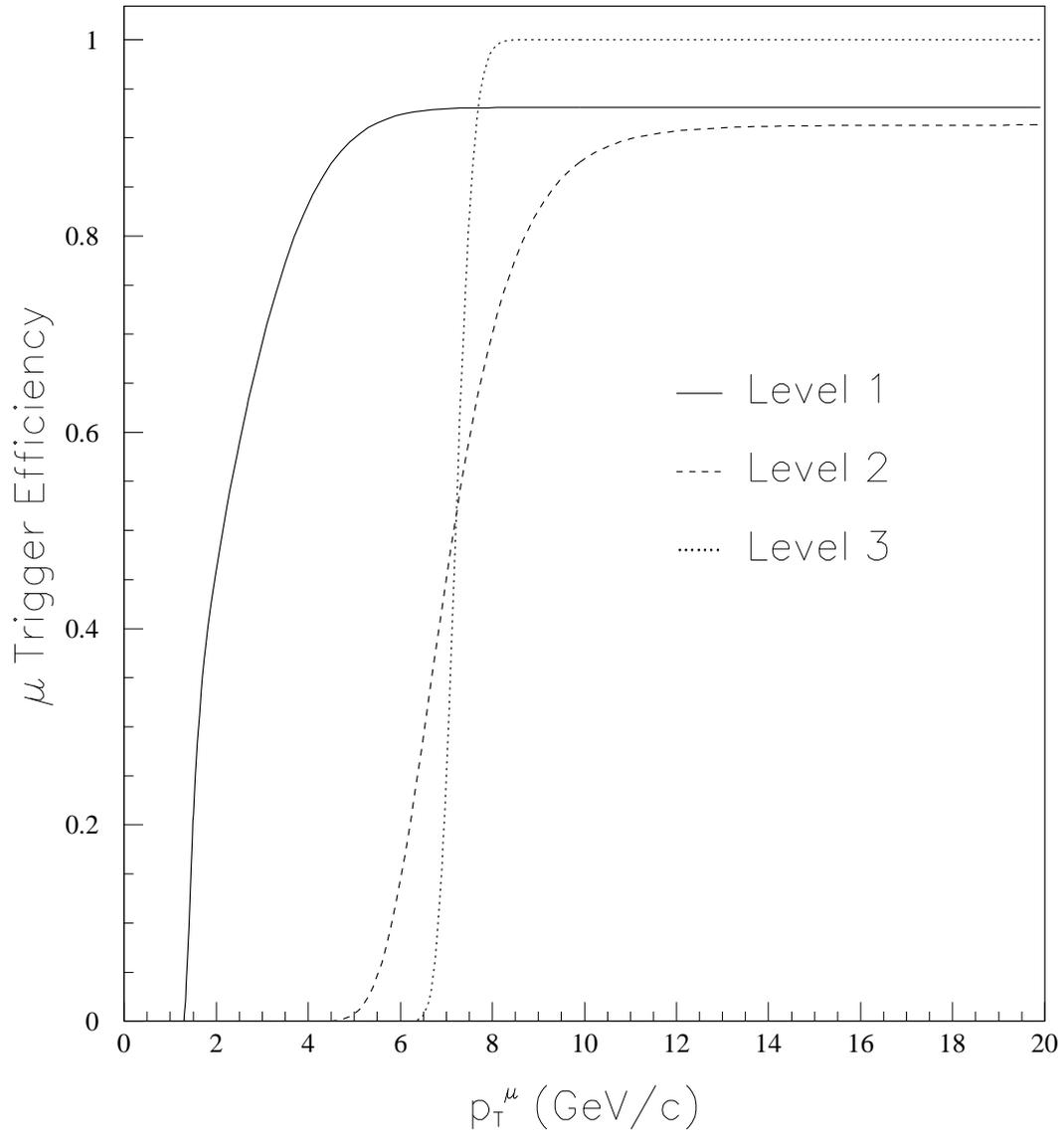}
\caption{The trigger efficiency curves for the 3 levels of the
trigger system.  The trigger efficiency is the product of the three
curves, convoluted with the $\mu~\pt$ spectrum.}
\label{fig-trigger_efficiency}
\end{figure}

  The trigger efficiency is measured using independently
triggered samples for each level of the system, where the efficiency
is expressed as a function of the muon $\pt$.
Figure~\ref{fig-trigger_efficiency} shows the efficiency curves for
the 3 levels of the trigger system.  The efficiency curves are then
convoluted with the $\pt$ spectrum of the muons, to extract the
efficiency for a muon with $\pt >$ 9 GeV/$c$.  This convolution is
done independently for the differential $\Et$ cross section bins (see
table~\ref{table-trigger_efficiency}), since the muon $\pt$ spectrum
may depend upon the transverse momentum distribution of the $\bbar$
jet recoiling against the $b \rightarrow \mu$ decay.  For $\bbar$
jets with $\Et >$ 10 GeV, the combined L1, L2, and L3 trigger
efficiency is measured to be 83.0 $\pm$ 1.7 \%.

\par The vertex requirement, $\mid z_0 \mid < 30$ cm, is studied in a
minimum bias trigger dataset, comparing the vertex distribution to
the predicted shape, including the measured  longitudinal
distribution of the proton and anti-proton bunches and the effects of
the accelerator $\beta$ function~\cite{Badgett_thesis}.  The
efficiency is found to be 74.2 $\pm$ 2.1 \%, where the uncertainty
comes from uncertainty in the measured beam longitudinal
distributions and $\beta$ function.

\par  The track finding efficiency in the CTC is a function of the
density of charged particles.  By embedding Monte Carlo simulated
track hits into data samples, we  quantify the probability of finding
the Monte Carlo simulated track  as a function of the relative
density of CTC hits.  The quantified probability is convoluted with
the hit density distribution  for the muon sample.  The track finding
efficiency is measured to be 96 $\pm$ 1.7 \%, where the uncertainty
represents the change in the result using different parametrizations
of the probability curve vs hit density.

\par    The combined $\chi^2$ matching efficiency is measured in a
$J/\psi \rightarrow \mu^+ \mu^-$ sample identified by tracking and
mass requirements and is found to be  $98.7 \pm 0.2$ \%.  The muon
segment reconstruction efficiency is found to  $98.1 \pm 0.3$ \%,
resulting in a combined efficiency of $96.8 \pm 0.4$ \%.

\par  The track finding efficiency in the SVX is studied in the 9
GeV/$c$ muon  sample, requiring the CTC track to extrapolate to a
good SVX fiducial region.  The efficiency is found to be 90 $\pm$
1\%, where the uncertainty is the statistical error only.

\begin{table}[tb]
\begin{center}
\begin{tabular}{|c|c|}       \hline
$\Et$ bin & Trigger Efficiency \\ \hline \hline
10 --- 15 GeV & 82.6 $\pm$ 1.7 \% \\	
15 --- 20 GeV & 83.0 $\pm$ 1.7 \% \\
20 --- 25 GeV & 83.4 $\pm$ 1.7 \% \\
25 --- 30 GeV & 83.6 $\pm$ 1.7 \% \\
30 --- 40 GeV & 83.8 $\pm$ 1.7 \% \\
40 --- 50 GeV & 83.9 $\pm$ 1.7 \% \\
50 --- 100 GeV & 83.7 $\pm$ 1.7 \% \\ \hline
All $\Et$ & 83.0 $\pm$ 1.7 \% \\
\hline
\end{tabular}
\end{center}
\caption{The muon trigger efficiency for each jet $\Et$ bin.  A
common 2\% uncertainty is assigned to each bin.}
\label{table-trigger_efficiency}
\end{table}

\begin{table}[tb]
\begin{center}
\begin{tabular}{|c|c|}       \hline
Geometric Acceptance & 53.0 $\pm$ 0.3 \% \\ \hline 	
CTC Track Finding  & 96.0 $\pm$ 1.7 \% \\		
Matching Efficiency & 96.8 $\pm$ 0.4 \% \\		
Z Vertex Requirements & 74.2 $\pm$ 2.1 \% \\		
SVX Track Finding & 90 $\pm$ 1 \% \\ \hline		
Combined Acceptance  & \\
and Efficiency & 32.9 $\pm$ 1.1 \% \\ \hline 		
\end{tabular}
\end{center}
\caption{Summary of muon acceptance and efficiency numbers.  The
trigger efficiency is applied on a bin by bin basis for the jet $\Et$
measurement.}
\label{table-efficiencies}
\end{table}

\subsection{$\bbar$ Jet Requirements}

\par  The $\bbar$ jet acceptance combines the fiducial acceptance of
the SVX and the CTC, the track reconstruction efficiency, and
fragmentation effects and the $\Delta$R separation requirement.
These tracking and $\Delta$R effects are studied separately, with a
full simulation used for the combination of the track requirements
and fiducial acceptance,  while a MNR based $\mu-\bbar$ model  is
used for the  $\Delta$R acceptance.  The ${\overline b}$ jet
acceptance is calculated separately as a function of the jet $\Et$
and azimuthal opening angle between the muon and the jet.

\par Monte Carlo samples for $b$ and $c$ quarks are produced using
ISAJET version 6.43~\cite{ISAJET}.  The CLEO Monte Carlo
program~\cite{CLEOMC} is used to model the decay of $B$ hadrons.  $b$
quarks produced using the HERWIG Monte Carlo~\cite{herwig} and PYTHIA
Monte Carlo~\cite{pythia} programs are also used for systematic
studies.  The ISAJET and PYTHIA samples used the Peterson form as the
fragmentation model, with $\epsilon =$ 0.006 $\pm$ 0.002.   While
none of these generators use a NLO calculation of $b$ production, the
$\eta$ distribution of the quarks agrees well with the NLO
calculation. For tracking efficiency studies, events with a muon with
$\pt >$ 8 GeV are passed through the full CDF simulation and
reconstruction package.  The simulation used an average $b$ lifetime
of $c\tau$ = 420 $\mu$m~\cite{lifetime_choice}.

\par The track acceptance represents the fraction of $\bbar$ quarks,
$\Et >$ 10 GeV, $|\eta| <$ 1.5 which produce jets with at least 2
good tracks inside a cone of 0.4 around the jet axis, where there is
also a $b$ quark which decays to a muon with $\pt >$ 9 GeV within the
CMU-CMP acceptance.    The average track acceptance for the
${\overline b}$ is 51.4 $\pm$ 0.8\%.  It ranges from 45.7 $\pm$ 1.1\%
(statistical error only) for 10 $<$ $\Et$ $<$ 15 GeV to 64.8 $\pm$
2.6\% for 50 $<$ $\Et$  $<$ 100 GeV.

\begin{table*}[t]
\begin{center}
\begin{tabular}{|c|c|c|}       \hline
$\Et$ Range & Track Acceptance & $\Delta$R Acceptance \\ \hline\hline
10 --- 15 & 45.7 $\pm$ 1.1 $\pm$ 2.3 \%
& 86.9 $\pm$ 1.0 $^{+1.4}_{-1.6}$ \% \\
15 --- 20 & 55.9 $\pm$ 1.7 $\pm$ 2.8 \%
& 88.2 $\pm$ 1.5 $^{+1.7}_{-1.9}$ \% \\
20 --- 25 & 58.1 $\pm$ 2.5 $\pm$ 2.9 \%
& 88.3 $\pm$ 2.0 $^{+2.2}_{-2.0}$ \% \\
25 --- 30 & 61.3 $\pm$ 3.5 $\pm$ 3.1 \%
& 88.3 $\pm$ 2.3 $^{+3.0}_{-3.5}$ \% \\
30 --- 40 & 61.7 $\pm$ 3.8 $\pm$ 3.1 \%
& 87.9 $\pm$ 3.4 $^{+3.6}_{-5.4}$ \% \\
40 --- 50 & 64.8 $\pm$ 2.6 $\pm$ 3.2 \%
& 87.1 $\pm$ 3.5 $^{+4.2}_{-5.1}$ \% \\
50 --- 100 & 65.0 $\pm$ 2.6 $\pm$ 3.3 \%
& 85.5 $\pm$ 3.7 $^{+5.2}_{-1.9}$ \% \\
\hline \hline
$\delta\phi$ Range (radians) &  & \\ \hline\hline
0---$\frac{\pi}{8}$  & 46.3 $\pm$ 1.4 $\pm$ 2.6 \%
& 6.9 $\pm$ 0.03 $^{+0.3}_{-0.2}$ \%\\
$\frac{\pi}{8}$ --- $\frac{\pi}{4}$
& 47.3 $\pm$ 1.4 $\pm$ 2.6 \%  & 20.8 $\pm$ 0.2 $^{+2.1}_{-0.3}$ \%\\
$\frac{\pi}{4}$ --- $\frac{3\pi}{8}$
& 51.4 $\pm$ 0.8 $\pm$ 2.6 \%  & 74.7 $\pm$ 0.9 $^{+6.0}_{-0.0}$ \%\\
$\frac{3\pi}{8}$ --- $\pi$
& 51.4 $\pm$ 0.8 $\pm$ 2.6 \%  & 100 \% \\ \hline
\end{tabular}
\end{center}
\caption{$\mu - \bbar$ track and $\Delta$R acceptance as a function
of jet $\Et$ and  $\delta\phi$ (statistical and systematic
uncertainties).  There is a common (relative) systematic uncertainty
of 5\% in the tracking efficiency.  For $\delta\phi >$ 1 radian, the
$\Delta$R acceptance is 100\% by definition.}
\label{table-bbar_jet_acceptance}
\end{table*}

\par  We have compared the values for the ${\overline b}$ track
acceptance from ISAJET samples to the acceptance from HERWIG samples.
The acceptance agrees within the statistical error in the samples as
a function of $\Et$, differing at the 5\% level.  We include this
variation as an additional systematic uncertainty on the track
acceptance.  Comparisons of inclusive jet track acceptances from an
ISAJET sample and from data show reasonable agreement.

\par For the calculation of the $\Delta$R acceptance, we have used a
model based on the MNR calculation~\cite{MNR}.  This calculation can
be used to give exact ${\cal O}(\alpha_s^3)$ results in situations
where kinematical cuts have been applied at the parton level.  We
have made additions to the calculation to model the $\mu - \bbar$
differential cross sections.

\par The MNR calculation~\cite{MNR} produces the vectors $p^b$,
$p^{\bbar}$, and $p^{gluon}$ with appropriate weights.  We include
additional weighting for the following:

\begin{itemize}
\item Probability of $\pt^{\mu} >$ 9 GeV/c for given $\pt^b$, $\muprob$
\item Probability of $\Et^{\bbar}$ jet in a given $\Et$ bin for given
$\pt^{\bbar}$, $\etprob$
\end{itemize}

\par
\begin{figure}
\epsfysize=7in
\epsffile[0 90 594 684]{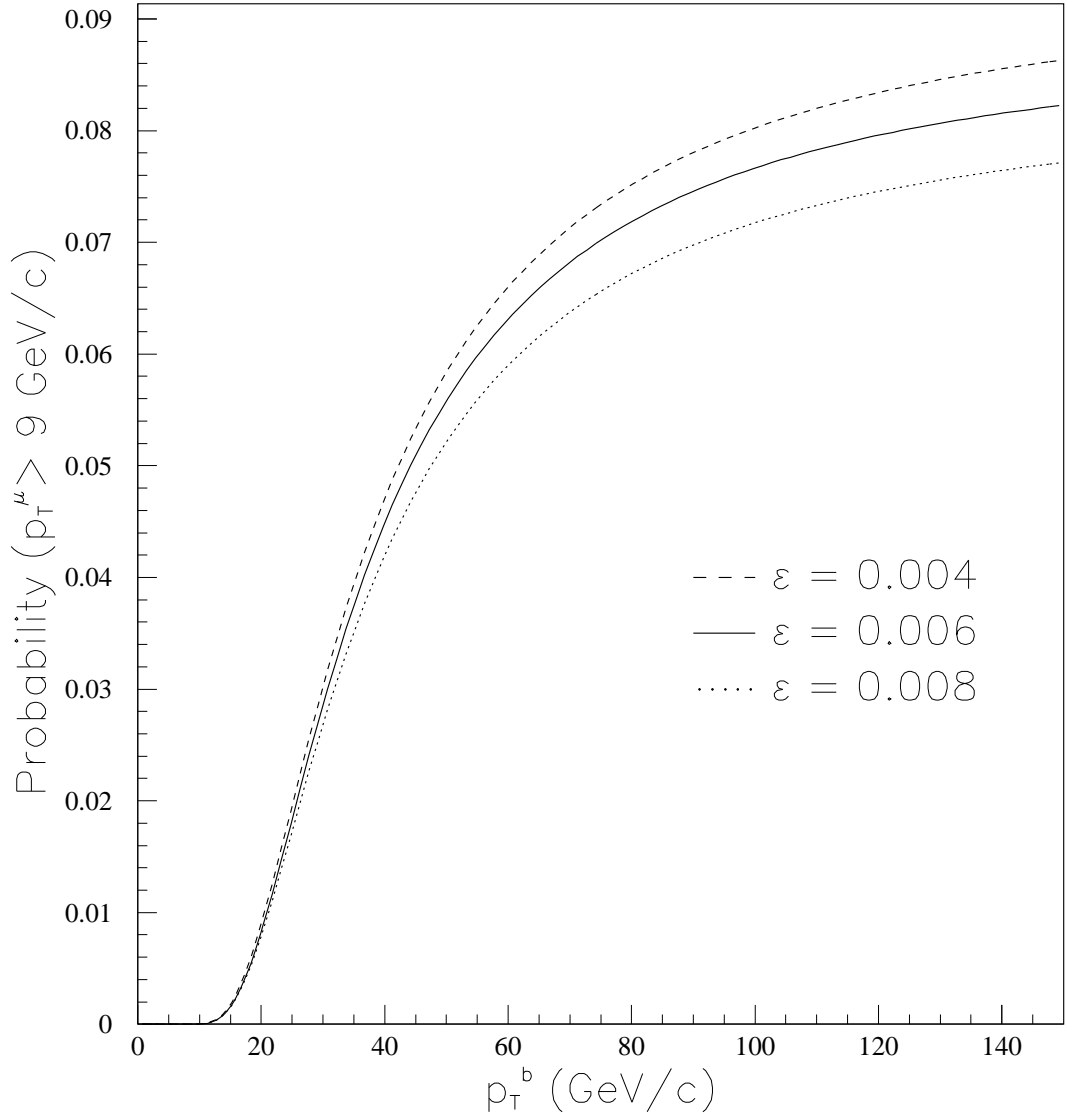}
\caption{The probability of a $b\rightarrow\mu$ decay, with
$\pt^{\mu} >$ 9  GeV/c, as a function of $\pt^b$.  We have included
the branching fraction $B(b\rightarrow\mu) = 0.103$.  The curves
represent three choices of the Peterson $\epsilon$ parameter used in
the fragmentation process.}
\label{fig-pt_mu_probability}
\end{figure}

 $\muprob$ is defined as the fraction of $b$ quarks, with given
$\pt^b$, which decay into muons with $\pt^{\mu} >$ 9 GeV/c.  We use
the $b\rightarrow\mu$ Monte Carlo generator described above to derive
this function, using B$(b\rightarrow\mu)$ = 0.103 $\pm$
0.005~\cite{PDG}.  Since the probability is defined as a function of
$\pt^b$, the exact shape of the $\pt^b$ distribution does not enter
into the result.  Figure~\ref{fig-pt_mu_probability} shows the value
$\muprob$ as a function of $\pt^b$.  The three curves are for
different values of the Peterson $\epsilon$ parameter used in the
fragmentation model.  In addition to this probability weighting, we
also smear the $b$ quark direction in pseudo-rapidity and azimuth.
The smearing is based on the results from the $b\rightarrow\mu$ Monte
Carlo generator.

\par  $\etprob$ is defined as the probability that a $\bbar$ quark,
with given $\pt^{\bbar}$, would produce a jet with given
$\Et^{\bbar}$.  Using the methods outlined in
section~\ref{section-results}, we have a binned probability
distribution in $\Et$ for each $\pt^{\bbar}$.  Since the measured jet
$\Et$ integrates over a range in pseudo-rapidity and azimuth (a cone
of radius 0.4), we approximate this clustering effect by clustering
partons (adding the $\bbar$  and the gluon momenta vectorially)
within the same cone size.  For the rest of this paper, when we
discuss the $\Et^{\bbar}$ or $\pt^{\bbar}$ theory distributions, it
means the clustered partons $\Et$ or $\pt$.

\par We use a renormalization and factorization scale $\mu_0 =
\sqrt{m_b^2 + (\pt^{b 2} + \pt^{\bbar 2})/2}$, MRSA structure
functions~\cite{MRSA}, and $m_b$ = 4.75 GeV/$c^2$.  Applying the
additional weights and the appropriate kinematical cuts ($\mid
\eta^{\mu} \mid <$ 0.6 and $\mid \eta^{\bbar} \mid <$ 1.5), we obtain
the calculated d$\sigma$/d$\Et^{\bbar}$ and
d$\sigma$/d$\delta\phi(\mu-\bbar)$ distributions.  We create the same
distributions with the requirement that the muon and $\bbar$ be
separated by $\Delta$R $>$ 1 and do a bin by bin comparison of the
calculated cross sections to define the $\Delta$R acceptance.  We
have varied the renormalization scale, $b$ quark mass, and parton
distribution functions used in the MNR calculation to estimate the
systematic uncertainties in the $\Delta$R acceptance.
Table~\ref{table-bbar_jet_acceptance} shows the bin by bin values
used in the differential cross section measurements.

\section{Cross Section Results}
\label{section-results}

\par The cross section results are presented as $\mu - \bbar$ cross
sections.  Since we have not specifically done flavor identification,
there is an additional factor of $1/2$ in the calculation of the
cross sections.  For the semi-differential measurements, we do an
independent fit of the log$_{10}(\pjet)$ distribution and then
correct for the acceptance in each $\Et$ or $\delta\phi$ bin.  With
the number of $\bbar$ jets from table~\ref{table-fit_results}, the
bin by bin trigger efficiencies from
table~\ref{table-trigger_efficiency}, the combined muon acceptance
and efficiency from table~\ref{table-efficiencies}, and the $\bbar$
track and $\Delta$R acceptances from
table~\ref{table-bbar_jet_acceptance}, we calculate the the cross
section in each $\Et$ and $\delta\phi$ bin considered. The sum of the
7 $\Et$ bins is 614.4 $\pm$ 63.0 pb and the sum of the 8 $\delta\phi$
bins is 633.0 $\pm$ 70.6 pb.  The results are summarized in
table~\ref{table-final_results}.

\begin{table}[tb]
\begin{center}
\begin{tabular}{|c||c|}       \hline
$\Et$ Range & Cross Section (pb) \\  \hline\hline
10 --- 15 & 168.1$^{+15.8}_{-15.8}$ \\
15 --- 20 & 152.2$^{+12.7}_{-12.8}$ \\
20 --- 25 & 106.6$^{+10.5}_{-10.5}$ \\
25 --- 30 & 61.93$^{+7.79}_{-7.79}$ \\
30 --- 40 & 72.53$^{+8.96}_{-9.43}$ \\
40 --- 50 & 29.81$^{+4.60}_{-4.68}$ \\
50 --- 100 & 23.13$^{+4.38}_{-4.23}$ \\  \hline \hline
$\delta\phi$ Range &  \\ \hline\hline
0---$\frac{\pi}{8}$ & 18.36$^{+24.38}_{-18.36}$ \\
$\frac{\pi}{8}$ --- $\frac{\pi}{4}$ & 30.86$^{+18.63}_{-10.91}$ \\
$\frac{\pi}{4}$ --- $\frac{3\pi}{8}$ & 17.30$^{+4.60}_{-4.21}$ \\
$\frac{3\pi}{8}$ --- $\frac{\pi}{2}$ & 18.48 $\pm$ 3.76 \\
$\frac{\pi}{2}$ --- $\frac{5\pi}{8}$ & 24.81 $\pm$ 4.37 \\
$\frac{5\pi}{8}$ --- $\frac{3\pi}{4}$ & 37.81 $\pm$ 5.91 \\
$\frac{3\pi}{4}$ --- $\frac{7\pi}{8}$ & 108.9 $\pm$ 9.93 \\
$\frac{7\pi}{8}$ --- $\pi$ & 376.4 $\pm$ 17.7 \\
\hline
\end{tabular}
\end{center}
\caption{$\mu - \bbar$ cross sections as a function of jet $\Et$ and
$\delta\phi$ between the muon and $\bbar$ jet.  We have not included
a common systematic uncertainty of 9.3\% in the results. Physics
backgrounds have not been subtracted at this stage.}
\label{table-final_results}
\end{table}

\subsection{Physics Backgrounds}

\par There are backgrounds which need to be included before comparing
to theoretical predictions on $b - \bbar$ production, since there are
additional sources of $\mu - \bbar$ production.  Specifically, the
decay products of light mesons ($\pi$, $K$) produced in association
with $b - \bbar$ pairs or heavy particles (e.g, the $Z\deg$ boson,
top quark production) can give a similar signature.

\par A contribution to the sample occurs when the identified muon is
not coming from a $b$ quark decay but instead from the decay of a
light meson ($\pi$ or $K$) or charm quark.  In the inclusive muon
sample, the $b$ fraction is measured to be approximately
40\%~\cite{TOPPRD}, with a charm fraction of approximately 20\% and
the remaining 40\% from the decay of light mesons.   Since jets from
gluons are the dominant production process in this jet $\Et$ range,
we assume that the light mesons come predominantly from gluon jets.
With the further assumption that  the gluon splitting to $b\bbar$
probability is approximately 1.5\%~\cite{gluon_splittin}, we estimate
that in 0.6\% ($0.015 \times 0.4$) of the muon events we correctly
identify the $\bbar$ but the muon is from a light meson decay.    The
case where the identified muon comes from the decay of a charm
particle can be estimated in a similar manner.  With the same
assumptions about the gluon splitting to heavy quark probability
(1.5\%), a measured charm fraction of 20\%, and  that approximately
75\% of charm quarks are produced via gluon splitting, we estimate
that in 0.2\% ($0.015 \times 0.75 \times 0.2$) of the muon events we
correctly identify the $\bbar$ but the muon is from a charm particle
decay.

\par With an identified fraction of 40\% $b$ muons and 50\% of the
produced $b$'s from gluon splitting~\cite{gluon_splittin}, in 20\% of
the muon events we correctly identify the $\bbar$ and the muon from
the $b$ decay.   Combining these calculations yields a fractional
background in the $\mu - \bbar$ cross section of 0.04 ($ = (0.006 +
0.002) / 0.20$). We assume that this  background has the same shape
as the signal and reduce the cross sections by a constant 4.0 $\pm$
2.0\% (the uncertainty is taken as half the change).

\par  We have used the PYTHIA Monte Carlo program to generate $Z\deg
\rightarrow b \bbar$ events, and the CLEO Monte Carlo program for the
decay of the resulting $B$ hadrons.  We normalize the production
cross section to measured CDF cross section of $Z\deg \rightarrow e^+
e^-$~\cite{PDG,my_result}, and apply the same $\mu$ and $\bbar$ jet
requirements as presented in section~\ref{section-dataset}.  The
predicted cross section remaining after these requirements is 3.6
$\pm$ 0.28 pb, where the uncertainty includes the relative
normalization to the dielectron decay mode, the $b \rightarrow \mu$
branching fraction, and acceptance uncertainties.
Table~\ref{table-z_b_bbar} shows the contributions from this process
in the same $\Et$ and $\delta\phi$ bins as in
table~\ref{table-final_results}.

\begin{table}[tb]
\begin{center}
\begin{tabular}{|c|c|}       \hline
$\Et$ Range &  Cross Section (pb) \\
 & Statistical Uncertainty only \\ \hline \hline
10 --- 15 & 0.43 $\pm$ 0.06 \\
15 --- 20 & 0.75 $\pm$ 0.08 \\
20 --- 25 & 0.82 $\pm$ 0.09 \\
25 --- 30 & 0.60 $\pm$ 0.07 \\
30 --- 40 & 0.87 $\pm$ 0.10 \\
40 --- 50 & 0.12 $\pm$ 0.02 \\
50 --- 100 & 0.015 $\pm$ 0.008 \\
\hline \hline
$\delta\phi$ Range &  \\ \hline\hline
0---$\frac{\pi}{8}$ & 0 \\
$\frac{\pi}{8}$ --- $\frac{\pi}{4}$ & 0 \\
$\frac{\pi}{4}$ --- $\frac{3\pi}{8}$ & 0.015 $\pm$ 0.014 \\
$\frac{3\pi}{8}$ --- $\frac{\pi}{2}$ & 0.031 $\pm$ 0.012 \\
$\frac{\pi}{2}$ --- $\frac{5\pi}{8}$ & 0.036 $\pm$ 0.013 \\
$\frac{5\pi}{8}$ --- $\frac{3\pi}{4}$ & 0.11 $\pm$ 0.024 \\
$\frac{3\pi}{4}$ --- $\frac{7\pi}{8}$ & 0.53 $\pm$ 0.05 \\
$\frac{7\pi}{8}$ --- $\pi$ & 2.88 $\pm$ 0.12 \\ \hline
\end{tabular}
\end{center}
\caption{Contributions from $Z\deg \rightarrow \mu \bbar$ to the
cross section as a function jet $\Et$ and $\delta\phi$ between the
$\mu$ and $\bbar$ jet. There is an addition 8.0\% uncertainty in the
overall normalization.}
\label{table-z_b_bbar}
\end{table}

\par Top quark production and decay can also contribute to the $\mu -
\bbar$ cross sections.  The CDF measurement of the total top cross
section is $6.8^{+3.6}_{-2.4}$ pb~\cite{TOPPRL95}.  However, once we
account for branching fractions and acceptance criteria, the total
cross section from this process is  less than 1 pb and will not be
considered further.

\subsection{Jet Unsmearing Procedure}

\par The cross sections measured above depend upon the selection of
jets with $\Et >$ 10 GeV, and in the case of the d$\sigma$/d$\Et$
distribution, depend upon the binning of the distribution.  Jets
coming from $\bbar$ quarks with transverse momentum $\pt^{\bbar}$
will contribute to more than one bin in the measured distribution,
due to the combined effects of calorimeter energy response,
calorimeter energy resolution, and quark fragmentation.  An
unsmearing procedure has been developed at CDF to account for these
effects.

\par We use Monte Carlo produced samples to define the expected jet
$\Et$ response distribution for a given quark $\pt$.  An iterative
procedure is used to correct the measured cross sections.  The quark
$\pt$ distribution is described by a smooth function and smeared with
the simulation derived $\Et$ response functions.  The input
distribution is adjusted until the smeared distribution matches the
measured distribution.  We then perform a simultaneous unfolding of
the measured jet $\Et$ spectrum to the parton $\pt$ spectrum to
account for energy loss and resolution.  This unfolding corrects both
the cross section and $\Et$ ($\pt$) axes.

\subsubsection{Response Functions}

\par The calorimeter single particle response in the range 0.5 to 227
GeV has been determined from both test beam data and isolated tracks
from collider  data.  A Monte Carlo simulation incorporating the
calorimeter response and the ISAJET, HERWIG, and PYTHIA samples is
used to determine a response function for $\bbar$ jets in the $\Et$
range 5 to 150 GeV, including energy loss, resolution, and jet
finding efficiency effects.  For each $\pt$, the response function
represents the probability distribution for measuring a particular
value of $\Et$.  These response functions are convoluted with the
expected $\bbar ~ \pt$ distributions, creating an expected $\Et$
distribution.

\subsubsection{Unsmearing}

\par
\begin{figure}
\epsfysize=7in
\epsffile[0 90 594 684]{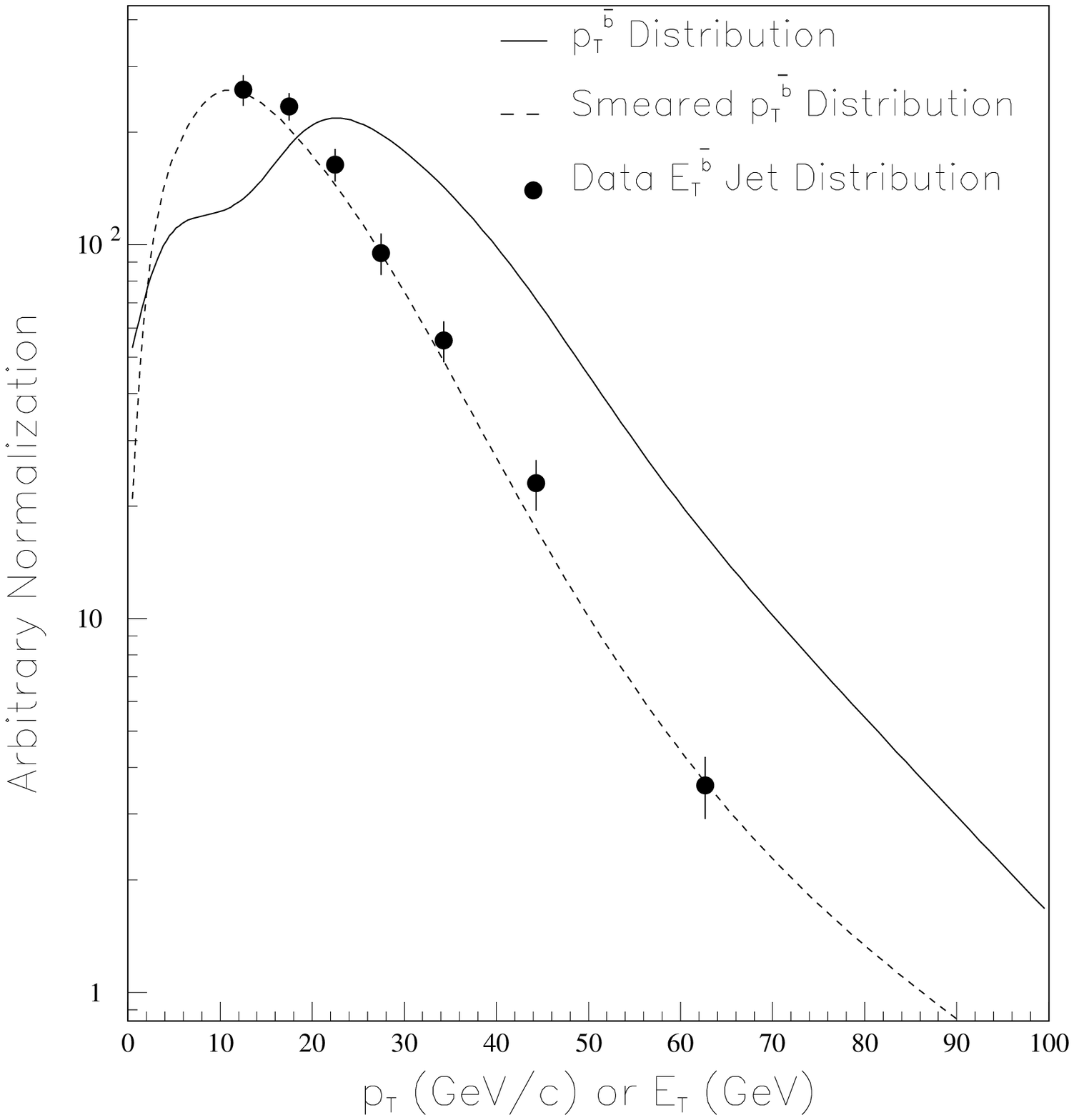}
\caption{The best match $\pt^{\overline b}$ distribution, overlayed
with the smeared distribution (dashed) and the data $\Et$
measurement.  The process is reversed to take the data $\Et$
distribution to a $\pt$ distribution.}
\label{fig-bbar_pt_distribution}
\end{figure}

  The input $\bbar$ distribution comes from the $\mu - \bbar$
model described in section~\ref{section-acceptance}, where we have
required a muon with $\pt>$ 9 GeV/c.  We have parametrized the
distribution with a multi-quadric function and varied a scale
parameter until the smeared distribution matches the measured
distribution.  Figure~\ref{fig-bbar_pt_distribution} shows the best
match $\bbar$ $\pt$ distribution, overlayed with the smeared
distribution.  Table~\ref{table-unsmeared} shows the unfolding
effects on the cross section and transverse momentum.  Note that the
unsmearing procedure introduces correlated systematic uncertainties
in the bins.

\begin{table*}[t]
\begin{center}
\begin{tabular}{|c|c|c||c|c|}       \hline
Jet $\Et$ Bin & Mean Jet $\Et$ & $\sigma$ (pb/GeV)
& Mean ${\overline b}~\pt$ & $\sigma$ (pb/GeV/c) \\
\hline \hline
10 ---  15 & 12.38 & 32.20  & 25.28 &  27.66  \\
15 ---  20 & 17.35 & 29.07  & 30.67 &  24.62  \\
20 ---  25 & 22.30 & 20.30  & 35.99 &  18.78  \\
25 ---  30 & 27.34 & 11.77  & 41.20 &  12.52  \\
30 ---  40 & 34.31 &  6.88  & 48.38 &   7.13  \\
40 ---  50 & 44.36 &  2.85  & 59.00 &   3.05  \\
50 --- 100 & 63.19 &  0.44  & 79.18 &   0.57  \\
\hline
\end{tabular}
\end{center}
\caption{Smeared and unsmeared means and cross sections for the 7
bins in the differential $\pt$ measurement.  The cross sections are
after background subraction and are presented here without
uncertainties.  Note that the unsmearing procedure introduces
correlated uncertainties in the bins.}
\label{table-unsmeared}
\end{table*}

\subsubsection{Systematic Uncertainties}

\par Systematic uncertainties in the smearing procedure arise from
uncertainties in the knowledge of the calorimeter energy scale, the
calorimeter resolution, the jet finding efficiency, the $\bbar$ quark
fragmentation, and the effects of the underlying event in defining
the jet energy.  The parameters in the smearing procedure are
adjusted to account for these uncertainties, the input distribution
is smeared, and the difference between the standard smeared
distribution and the new smeared distribution is used to estimate the
bin by bin systematic uncertainties.  The uncertainties are added in
quadrature to extract a total systematic uncertainty.
Table~\ref{table-all_systematics} contains the bin by bin systematic
uncertainties.

\begin{table*}[t]
\begin{center}
\begin{tabular}{|c|c|c|c|}       \hline
Variation & 10 --- 15 GeV $\Et$ & 15 --- 20 GeV $\Et$
& 20 --- 25 GeV $\Et$ \\
\hline \hline
 Energy Scale           &  + 7.2\% - 4.6\%
&  + 4.7\% - 3.5\%  &  + 9.1\% - 7.3\% \\
  Underlying event      &  + 0.2\% - 0.2\%
&  + 0.1\% - 0.2\%  &  + 0.2\% - 0.2\% \\
 Calorimeter Resolution &  + 4.4\% - 4.2\%
&  + 2.6\% - 2.5\%  &  + 4.1\% - 4.1\% \\
  Jet Finding           & $\pm$  2.6\%
& $\pm$  0.7\%  & $\pm$  1.0\% \\
  $b$ Fragmentation    & +  1.0\%  & -  4.0\%  & -  4.7\% \\
 \hline
  Total                 &  + 8.9\% - 6.7\%
&  + 5.4\% - 5.9\% &  +10.0\% - 9.6\%\\

\hline \hline
 & 25 --- 30 GeV $\Et$ & 30 --- 40 GeV $\Et$  & 40 --- 50 GeV $\Et$ \\
\hline \hline
 Energy Scale           &  +12.5\% -10.2\%
&  +16.5\% -13.4\%  &  +20.7\% -16.5\% \\
  Underlying event      &  + 0.3\% - 0.3\%
&  + 0.4\% - 0.4\%  &  + 0.4\% - 0.4\% \\
 Calorimeter Resolution &  + 3.5\% - 3.5\%
&  + 0.9\% - 0.9\%  &  + 4.5\% - 4.5\% \\
  Jet Finding           & $\pm$  0.3\%
& $\pm$  0.0\%  & $\pm$  0.0\% \\
  $b$ Fragmentation     & - 4.4\% & - 3.4\% & + 1.6\% \\
 \hline
  Total                 &  +13.0\% -11.6\%
&  +16.5\% -13.9\% &  +21.2\% -17.2\%\\
\hline \hline
  & 50 --- 100 GeV $\Et$ & & \\
\hline \hline
 Energy Scale           &  +27.8\% -21.3\% & & \\
  Underlying event      &  + 0.4\% - 0.4\% & & \\
 Calorimeter Resolution &  +12.7\% -12.7\% & & \\
  Jet Finding           & $\pm$  0.0\% & & \\
  $b$ Fragmentation     & + 1.0\% & & \\
 \hline
  Total                 & +30.6\% -24.8\% & & \\
\hline
\end{tabular}
\end{center}
\caption{Systematic uncertainties for each bin in the $\mu -
{\overline b}$ differential jet $\Et$ distribution.  There are bin to
bin correlations for each systematic variation.}
\label{table-all_systematics}
\end{table*}

\subsubsection{$\bbar$ Jet $\pt^{min}$ Definition}

\par  For future comparisons to theoretical predictions on overall
normalization, we need to define a $\pt^{min}$ threshold for the
recoiling $\bbar$ quark.  The standard definition is to take the
$\pt$ value where $>$90\% of all decays pass the kinematic cuts.  In
this case, we need to find the point where $>$ 90\% of all jets have
E$_T>$ 10 GeV.  We begin with the $\bbar$ $\pt$ spectrum shown in
figure~\ref{fig-bbar_pt_distribution} and apply the resolution
smearing to this distribution.  We weight each bin in the $\pt$
spectrum by the probability that a $\bbar$ quark with that $\pt$
would give a jet with E$_T>$ 10 GeV.  Integrating the resulting
weighted distribution gives a 90\% $\pt^{min}$ value of 20.7 GeV/c
for the $\bbar$ jet.

\begin{figure}
\epsfysize=7in
\epsffile[0 90 594 684]{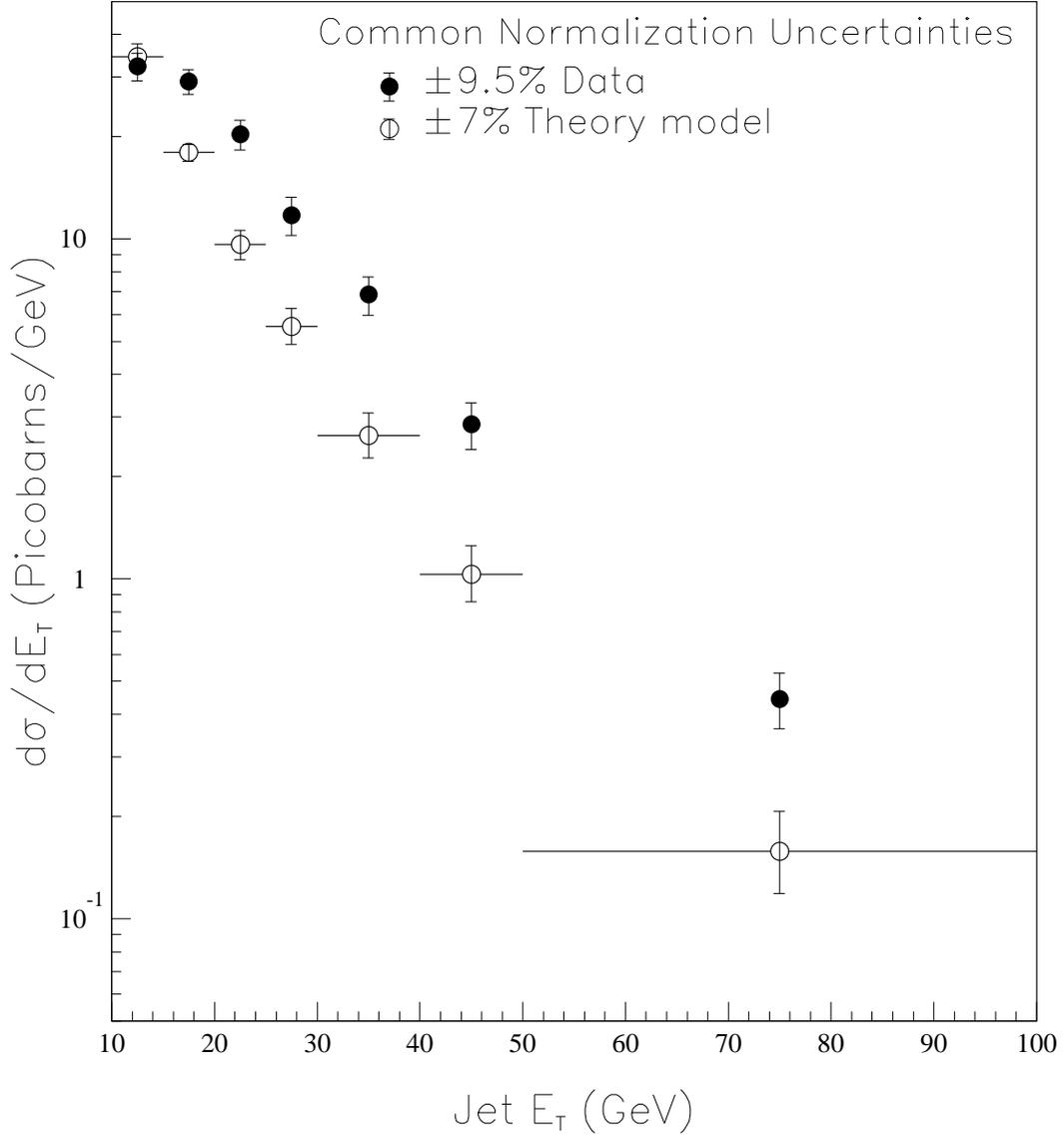}
\caption{The differential $\Et$ cross section, for
$\pt^{\mu}>~9$~GeV/c, $|\eta^{\mu}|<$~0.6, $\Et^{\bbar}>$~10~GeV,
$|\eta^{\bbar}|<$~1.5, compared to theoretical predictions.  The data
points have a common systematic of $\pm$ 9.5\%.  The common
uncertainty in the theory points comes from the muonic branching
fraction and fragmentation model.  The theory points do include
uncertainties from the smearing procedure.  }
\label{fig-et_comparison}
\end{figure}

\subsection{Comparison with NLO QCD}

\par  In figure~\ref{fig-et_comparison}, we show a comparison of the
differential jet $\Et$ cross section,

\begin{eqnarray}
\frac{{\rm d}\sigma}{{\rm dE}_{\rm T}^{\bbar}}(\pt^{\mu}> 9~{\rm GeV/c},
|\eta^b|<1,
|\eta^{\bbar}| < 1.5, \Et^{\bbar} > 10~{\rm GeV}) \nonumber
\end{eqnarray}

\noindent to a prediction from the $\mu - \bbar$ model discussed in
section~\ref{section-dataset}.  There is a 9.5\% common uncertainty
in the measured points, coming from the jet probability fit (5.8\%),
the $\bbar$ jet tracking efficiency (5\%), the muon acceptance and
identification efficiencies (3.9\%), the luminosity normalization
(3.6\%), and the remaining background subtraction (2\%).  This common
uncertainty is displayed separately. The uncertainty in the model
prediction represents the uncertainty from the muonic branching
fraction (5\%)~\cite{PDG}, the acceptance of the muon $\pt$ cut from
variations in the fragmentation model (5\%), which are common to all
points, and the uncertainties associated with  $\pt$ to $\Et$
smearing.  The data has an integral value of 586. $\pm$ 61.8 pb,
while the model predicts an integral value of 383.5 $\pm$ 5.9 pb.

\begin{figure}
\epsfysize=7in
\epsffile[0 90 594 684]{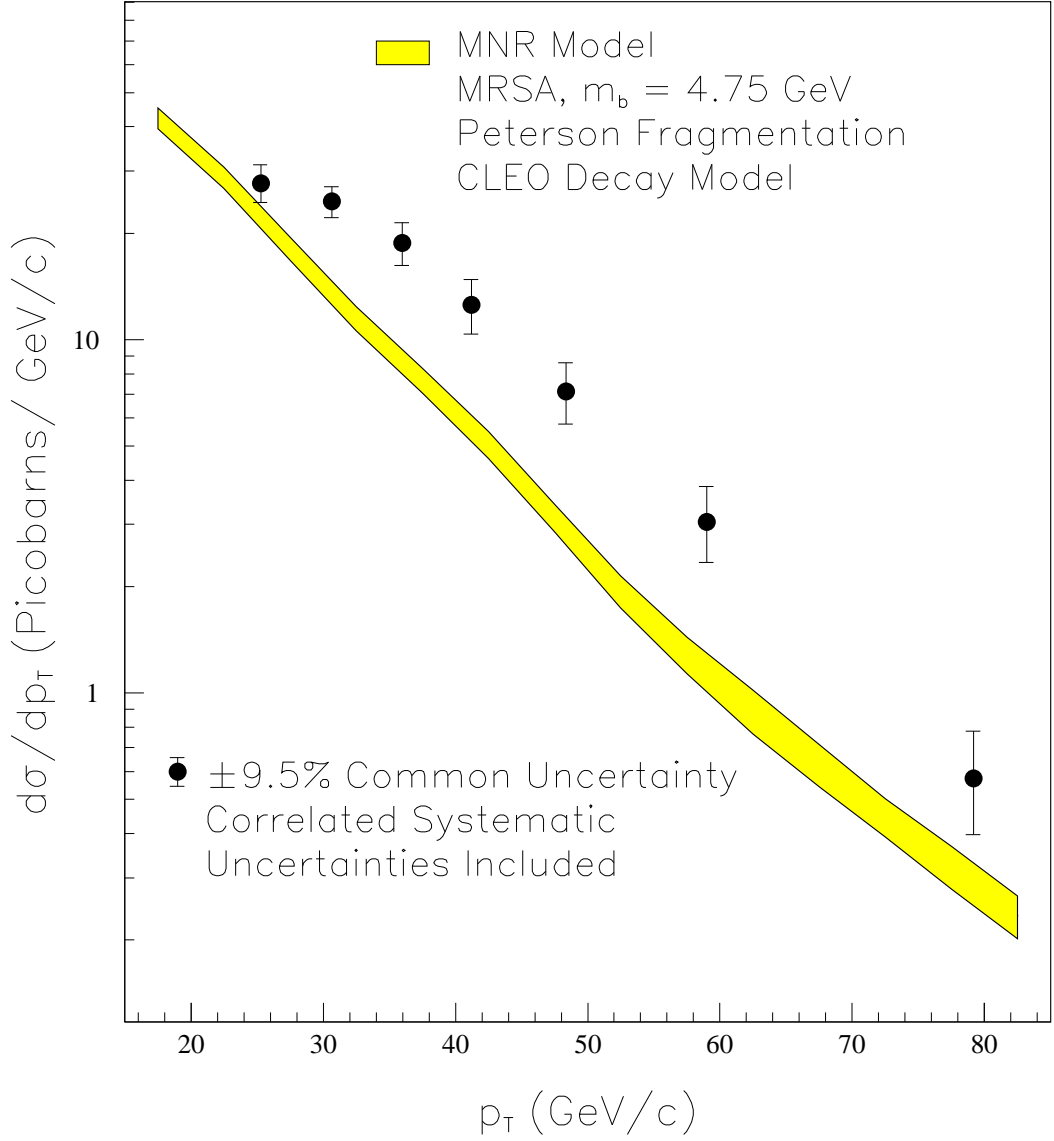}
\caption{The differential $p_t$ cross section, for
$\pt^{\mu}>~9$~GeV/c, $|\eta^{\mu}|<$~0.6, $|\eta^{\bbar}|<$~1.5,
compared to theoretical predictions.  The data points have a common
systematic of $\pm$ 9.5\% and there are correlated systematic
uncertainties.  The uncertainty in the theory curves comes from the
muonic branching fraction and fragmentation model.  }
\label{fig-pt_comparison}
\end{figure}

\par  In figure~\ref{fig-pt_comparison}, we show the unsmeared
differential jet $\pt$ cross section,

\begin{eqnarray}
\frac{{\rm d}\sigma}{{\rm d}\pt^{\bbar}}(\pt^{\mu}> 9~{\rm GeV/c},
|\eta^{\mu}| <0.6, |\eta^{\bbar}| < 1.5)  \nonumber
\end{eqnarray}

\noindent compared to the $\bbar~\pt$ prediction from the $\mu -
\bbar$ model, where we have included systematic uncertainties
associated with the resolution smearing on the measured points.
Again, the common normalization uncertainties are displayed
separately.

\begin{figure}
\epsfysize=7in
\epsffile[0 90 594 684]{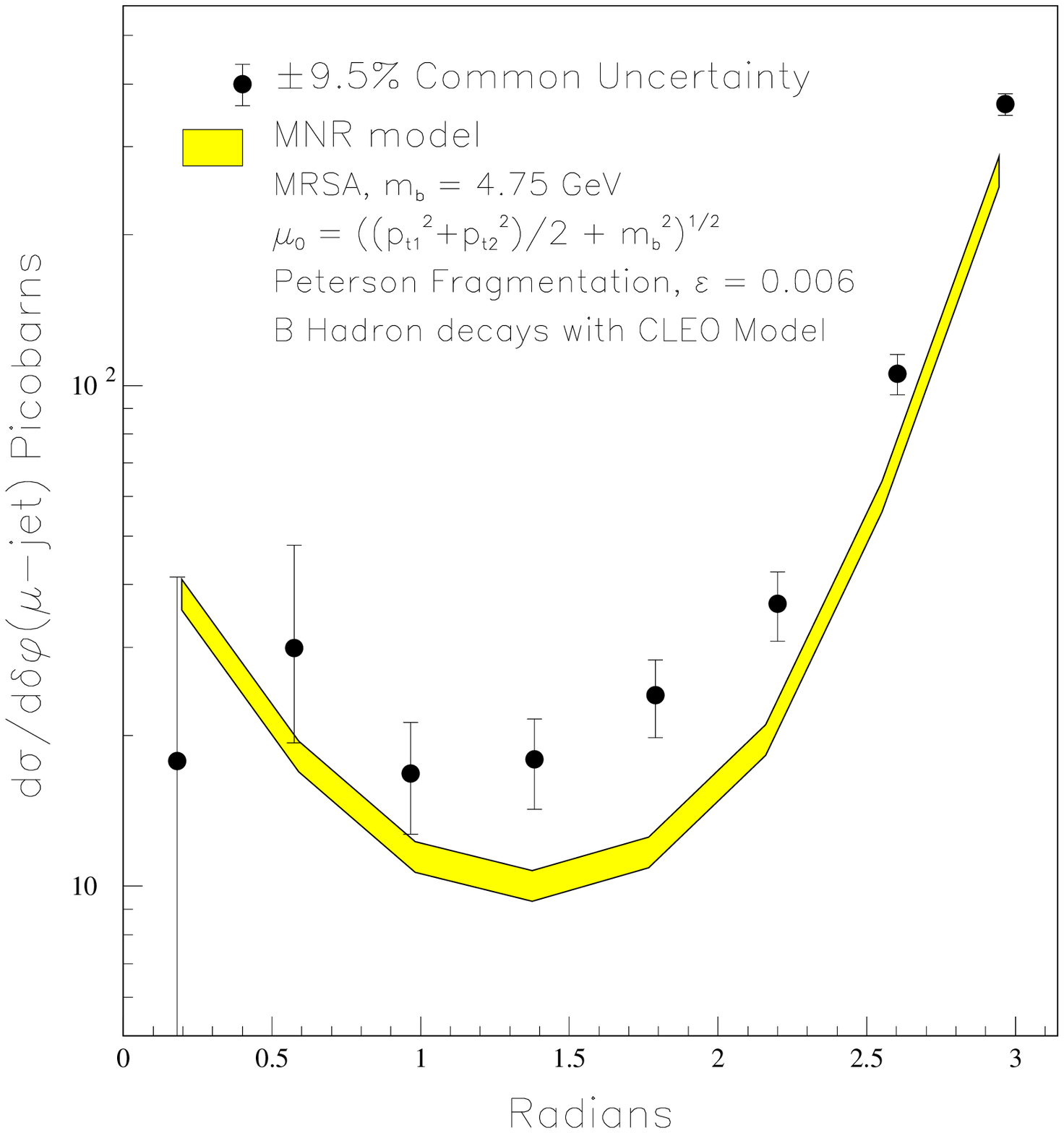}
\caption{The differential $\delta\phi$ cross sections, for
$\pt^{\mu}>~9$~GeV/c, $|\eta^{\mu}|<$~0.6, $\Et^{\bbar}>$~10~GeV,
$|\eta^{\bbar}|<$~1.5 compared to theoretical predictions.  The data
points have a common systematic of $\pm$ 9.5\%.  The uncertainty in
the theory curves comes from the muonic branching fraction and
fragmentation model.  }
\label{fig-phi_comparison}
\end{figure}

\par  In figure~\ref{fig-phi_comparison}, we show a comparison of the
differential $\delta\phi(\mu - \bbar)$ cross section,

\begin{eqnarray}
\frac{{\rm d}\sigma}{{\rm d}\delta\phi^{\mu-\bbar}}(\pt^{\mu}>9~{\rm GeV/c},
|\eta^{\mu}|<0.6,\Et^{\bbar}>10~{\rm GeV},|\eta^{\bbar}| < 1.5)
\nonumber
\end{eqnarray}

\noindent to the predictions from the $\mu - \bbar$ model.  The
uncertainty in the theoretical prediction represents the uncertainty
in the muonic branching fraction and fragmentation model only.

\par
\begin{figure}
\epsfysize=7in
\epsffile[0 90 594 684]{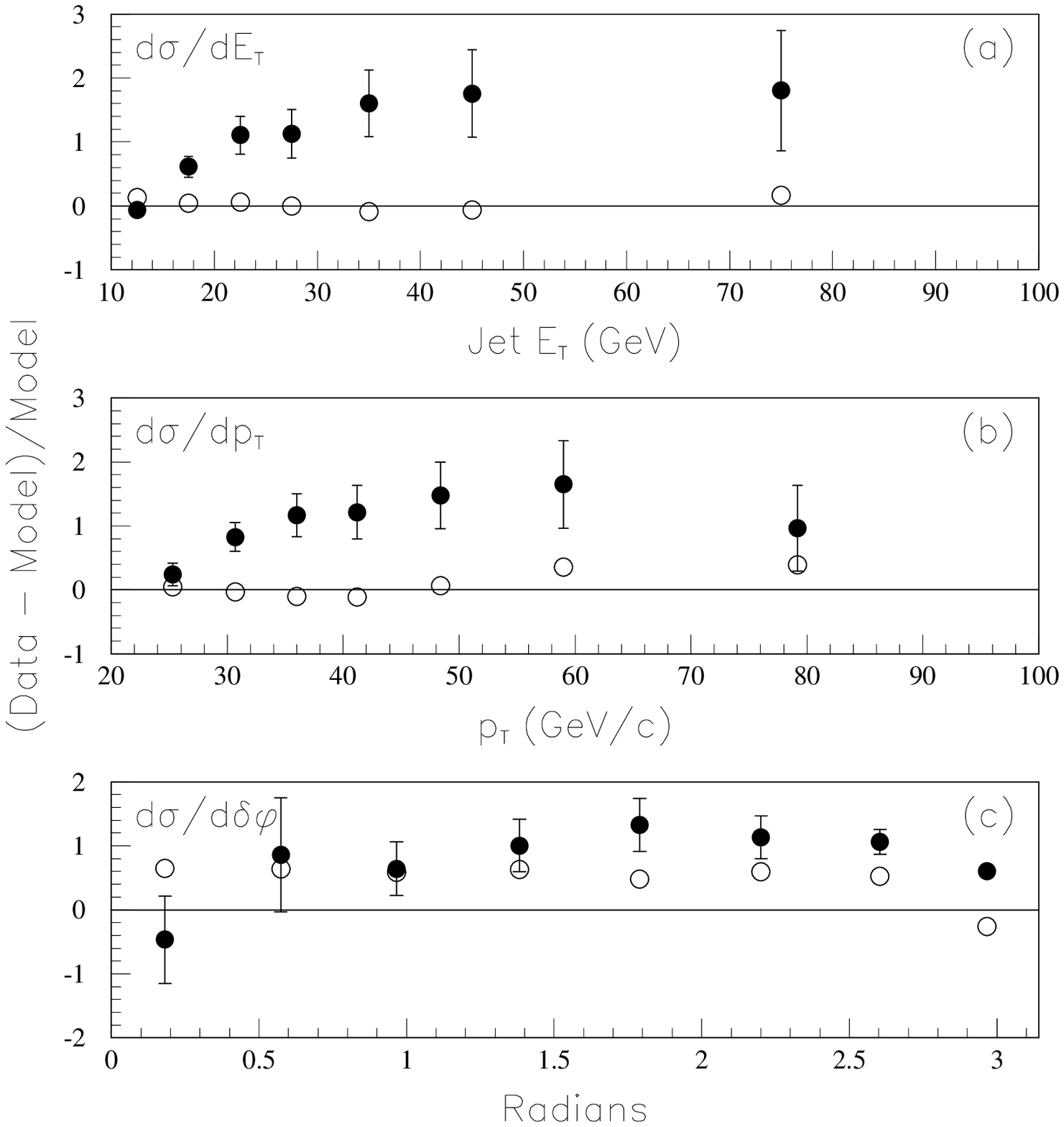}
\caption{For the (a) d$\sigma$/d$\Et$, (b) d$\sigma$/d$\pt$, and (c)
d$\sigma$/d$\delta\phi$ distributions, we plot the difference between
the data measurement (filled circles) and the model prediction,
scaled to the model prediction.  There is a common systematic
uncertainty of 9.5\% in all the points, which has not been included
in the error bar.  The open circles are the model prediction for
renormalization scale of $\mu_0/2$ minus the model prediction for
$\mu_0$, scaled to the model prediction for $\mu_0$.}
\label{fig-data_minus_qcd}
\end{figure}

 While we find qualitative agreement in shape between the
measured distributions and model predictions, there are some
differences.  To investigate in more detail, we present in
figure~\ref{fig-data_minus_qcd} the experimental results minus the
model prediction, scaled to the model prediction for the $\Et$,
$\pt$, and $\delta\phi$ distributions.   The $\Et (\pt$)
distributions have similar shapes for $\Et (\pt) >$ 20 GeV( 35
GeV/c), but different normalizations.  At lower values of $\Et$
($\pt$), the measurements and predictions are in agreement. The data
$\delta\phi$ distribution is somewhat broader than the model
predictions, with enhancement in the region $\pi/4$ to $3\pi/4$,  as
well as being at consistently higher values.  We have also shown how
the model prediction changes with change of the renormalization and
factorization scale, by plotting the prediction for scale $\mu_0/2$
minus the prediction for $\mu_0$, scaled to the prediction for
$\mu_0$.  The integral cross section increases by 7\%, with very
little change as a function of $\Et$ or $\pt$.  In the $\delta\phi$
distribution, the $\mu_0/2$ prediction is uniformly larger than the
$\mu_0$ prediction, except for the region $\delta\phi \approx \pi$.

\par  Recent work has shown that the addition of an intrinsic $\Kt$
kick to a next-to-leading order QCD calculation improves the
agreement between measurements and predictions for both direct photon
production~\cite{CTEQ_kt} and charm production~\cite{FMNR}.  We have
investigated the effects of additional intrinsic $\Kt$ in the $\mu -
\bbar$ model.  We use a Gaussian distribution with mean 0 and
adjustable width to model the magnitude of the kick, with a random
azimuthal direction.  With widths of 2 - 4 GeV/c, we find that the
dominant effects occur for $\delta\phi <$ 1 radian. The cross section
for $\delta\phi <$ 1 is predicted to change by approximately 7\% with
a width of 4 GeV/c. With the current statistical uncertainties at
small $\delta\phi$ (ranging from 25 - 100 \%), we are unable to
distinguish effects at that level.  Similarly, the dominant effect in
the $\pt^{\bbar}$ distribution occurs in regions where we have no
sensitivity ($\pt^{\bbar} <$ 20 GeV/c).  We conclude that the
addition of intrinsic $\Kt$ with width of 4 GeV/c does not account
for the difference between the model prediction and the measurement.

\section{Conclusions}
\label{section-conclusions}

\par  We have presented results on the semi-differential $\mu -
\bbar$ cross sections as a function of the $\bbar$ jet transverse
energy (d$\sigma$/d$\Et$), $\bbar$ transverse momentum
(d$\sigma$/d$\pt$), and the azimuthal opening angle between the muon
and the $\bbar$ jet (d$\sigma$/d$\delta\phi$).  These results are
based on precision track reconstruction in jets.  The effects of
detector response and resolution have been unfolded to translate the
results from $\bbar$ jets to $\bbar$ quarks.  We have compared these
results to a model based on a full NLO QCD calculation~\cite{MNR}.
We have investigated the effects an additional intrinsic $\Kt$ and
find that it cannot account for the difference between the
measurements and the model prediction.  Unlike previous CDF
measurements~\cite{cdf_bmu,CDF_Bmes}, a normalization change alone
does not account for the differences between this measurement and the
model prediction.

\par We thank the Fermilab staff and the technical staffs of the
participating institutions for their vital contributions.  This work
was supported by the U.S. Department of Energy and National Science
Foundation; the Italian Istituto Nazionale di Fisica Nucleare; the
Ministry of Education, Science and Culture of Japan; the Natural
Sciences and Engineering Research Council of Canada; the National
Science Council of the Republic of China; the A. P. Sloan Foundation;
and the Alexander von Humboldt-Stiftung.



\begin{thebibliography}{99}



\bibitem{cdf_bmu}
F. Abe, {\em et al.}, Phys. Rev. Lett., {\bf 71}, 2396 (1993).

\bibitem{D0-mu}
S. Abachi, {\em et al.}, Phys. Rev. Lett., {\bf 74}, 3548 (1995).

\bibitem{FMNR}
S. Frixione, {\em et al.}, Nucl. Phys. {\bf B 431}, 453 (1994).

\bibitem{CDF_Bmes}
F. Abe, {\em et al.}, FERMILAB-PUB-95/48-E, submitted to Phys. Rev. Lett.

\bibitem{snowmass_beta}
F. DeJongh, in the Proceedings of the Workshop on B Physics at Hadron
Colliders, ed. by P. McBride and C.S. Mishra, Snowmass, Co., June 1993.

\bibitem{coordinates}
The CDF coordinate system defines $Z$ along the proton-beam direction, $\theta$
as the polar angle, and $\phi$ as the azimuthal angle.  The pseudorapidity,
$\eta$, is defined as $\eta = - ln(tan(\theta/2)$.

\bibitem{CDF_emu}
F. Abe, {\em et al.}, Fermilab-PUB-94/131-E.


\bibitem{PDG}
L. Montanet, {\em et al.}, Particle Data Group, Phys. Rev. {\bf D50}, 1173
(1994).

\bibitem{jet_papers}
F. Abe, {\em et al.}, Phys. Rev. {\bf D45}, 1448 (1992);\\
F. Abe, {\em et al.}, Phys. Rev. {\bf D47}, 4857 (1993).


\bibitem{NIM_book}
F. Abe, {\em et al.}, Nucl. Instr. Meth.  {\bf A271}, 387 (1988) and
references therein.

\bibitem{SVXNIM}
D. Amidei, {\em et al.}, Nucl. Instr. Meth. {\bf A350}, 73 (1994).

\bibitem{CTCNIM}
F. Bedeschi, {\em et al.}, Nucl. Instr. Meth. {\bf A268}, 51 (1988).

\bibitem{CMUNIM}
G. Ascoli, {\em et al.}, Nucl. Instr. Meth. {\bf A268}, 33 (1988).

\bibitem{CMP}
J. Chapman, {\em et al.}, to be submitted to Nucl. Instr. Meth..

\bibitem{TOPPRD}
F. Abe, {\em et al.}, Phys. Rev. {\bf D50} 2966 (1994).

\bibitem{TRIGGERNIM}
D. Amidei, {\em et al.}, Nucl. Instr. Meth. {\bf A269}, 51 (1988).

\bibitem{CFTNIM}
G. W. Foster, {\em et al.}, Nucl. Instr. Meth. {\bf A269}, 93 (1988).


\bibitem{ALEPH}
D. Buskulic, {\em et al.}, Phys. Lett. {\bf B 313}, 535 (1993).

\bibitem{b_lifetime}
F. Abe, {\em et al.}, Phys. Rev. Lett. {\bf 71}, 3421 (1993), CDF presents
an error of 5.8\%.

\bibitem{lifetime_choice}
In the fall of 1993, the world average $B$ hadron lifetime was 1.4 psec, from
LEP and TeVatron experiments (see R. van Kooten, in {\em Results and
Perspectives in Particle Physics}, Proceedings of the 7th Rencontres de
Physique de la Vallee d'Aoste, La Thuile, Italy, 1993, ed. by M. Greco
(Editions Frontieres, Gif-sur-Yvette)).  Uncertainty in this quantity is
included as a systematic in the fitting procedure.


\bibitem{Badgett_thesis}
W. Badgett, Ph.D. thesis, University of Michigan, 1994 (unpublished).

\bibitem{MNR}
M. Mangano, P. Nason, and G. Ridolfi, Nucl. Phys. {\bf B373}, 295 (1992).
The FORTRAN code is available from the authors.

\bibitem{MRSD0}
A.D. Martin, R.G. Roberts, and W.J. Stirling, RAL-92-021 (1992).

\bibitem{Peterson}
C. Peterson, {\em et al.}, Phys. Rev. {\bf D 27}, 105 (1983).

\bibitem{fragmentation}
J. Chrin, Z. Phys.  {\bf C 36}, 165 (1987).

\bibitem{CLEOMC}
P. Avery, K. Read, and G. Trahern, Cornell Internal Note CSN-212, March 25,
1985 (unpublished).

\bibitem{ISAJET}
F. Paige and S.D. Protopopescu, BNL Report No. 38034, 1986 (unpublished).  We
used version 6.36.

\bibitem{herwig}
G. Marchesini and B.R. Webber, Nucl. Phys. {\bf B310}, (1988), 461;\\
G. Marchesini, {\em et al.}, Comput. Phys. Comm. {\bf 67} 465 (1992).

\bibitem{pythia}
H. Bengtsson and T. Sj$\ddot{{\rm o}}$strand, Comput. Phys. Commun. {\bf 46},
43 (1987).

\bibitem{MRSA}
A.D. Martin, W.J. Stirling, and R.G. Roberts, RAL-94-055, DTP/94/34 (1994).



\bibitem{my_result}
F. Abe, {\em et al.}, Phys. Rev. {\bf D44}, 29 (1991).

\bibitem{TOPPRL95}
F. Abe, {\em et al.}, Phys. Rev. Lett. {\bf 74}, 2626 (1995).

\bibitem{gluon_splittin}
These estimates are derived using the next-to-leading order calculations from
reference~\cite{MNR}; M. Seymour, Z. Phys. C {\bf 63} 99 (1995).

\bibitem{CTEQ_kt}
J. Huston, {\em et al.}, Phys. Rev. {\bf D51}, 6139 (1995).


\end{thebibliography}
\end{document}